\documentclass[aps,prl,superscriptaddress,twocolumn]{revtex4-2}
\usepackage{graphicx} 
\usepackage{subfigure}
\usepackage{multirow}
\usepackage{xspace}
\usepackage{fancyhdr}
\usepackage{amsmath,amssymb}
\usepackage{xcolor}

\usepackage{lineno}

\newcommand{\tc}{$T_{\text{c}}$\xspace}
\newcommand{\tdelta}{{$2\Delta$}\xspace}

\newcommand{\kv}{kV/cm\xspace}

\begin{document}
\title{Revealing novel aspects of light-matter coupling by terahertz two-dimensional coherent spectroscopy: the case of the amplitude mode in superconductors}
\author{Kota~Katsumi}
\email{kk5461@nyu.edu}
\affiliation{William H. Miller III , Department of Physics and Astronomy, The Johns Hopkins University, Baltimore, MD, 21218, USA} 
\author{Jacopo Fiore}
\affiliation{Department of Physics and ISC-CNR, Sapienza University of Rome, P.le A. Moro 5, 00185 Rome, Italy}
\author{Mattia Udina}
\affiliation{Department of Physics and ISC-CNR, Sapienza University of Rome, P.le A. Moro 5, 00185 Rome, Italy}
\author{Ralph~Romero III}
\affiliation{William H. Miller III , Department of Physics and Astronomy, The Johns Hopkins University, Baltimore, MD, 21218, USA} 
\author{David~Barbalas}
\affiliation{William H. Miller III , Department of Physics and Astronomy, The Johns Hopkins University, Baltimore, MD, 21218, USA} 
\author{John~Jesudasan}
\affiliation{Department of Condensed Matter and Material Science, Tata Institute of Fundamental Research, Homi Bhabha Rd., Colaba, Mumbai 400005, India}
\author{Pratap Raychaudhuri}
\affiliation{Department of Condensed Matter and Material Science, Tata Institute of Fundamental Research, Homi Bhabha Rd., Colaba, Mumbai 400005, India}
\author{Goetz Seibold}
\affiliation{Institut f\"{u}r Physik, BTU Cottbus-Senftenberg, P. O. Box 101344, 03013 Cottbus, Germany}
\author{Lara Benfatto}
\affiliation{Department of Physics and ISC-CNR, Sapienza University of Rome, P.le A. Moro 5, 00185 Rome, Italy}
\author{N.~P.~Armitage}
\affiliation{William H. Miller III , Department of Physics and Astronomy, The Johns Hopkins University, Baltimore, MD, 21218, USA} 

\begin{abstract}
	Recently developed terahertz (THz) two-dimensional coherent spectroscopy (2DCS) is a powerful technique to obtain materials information in a fashion qualitatively different from other spectroscopies. Here, we utilized THz 2DCS to investigate the THz nonlinear response of conventional superconductor NbN. Using broad-band THz pulses as light sources, we observed a third-order nonlinear signal whose spectral components are peaked at twice the superconducting gap energy \tdelta. With narrow-band THz pulses, a THz nonlinear signal was identified at the driving frequency $\Omega$ and exhibited a resonant enhancement at temperature when $\Omega = 2\Delta$.  General theoretical considerations show that such a resonance can only arise from a disorder-activated paramagnetic coupling between the light and the electronic current. This proves that the nonlinear THz response can access processes distinct from the diamagnetic Raman-like density fluctuations, which are believed to dominate the nonlinear response at optical frequencies in metals. Our numerical simulations reveal that even for a small amount of disorder, the $\Omega=2\Delta$ resonance is dominated by the superconducting amplitude mode over the entire investigated disorder range. This is in contrast to other resonances, whose amplitude-mode contribution depends on disorder.  Our findings demonstrate the unique ability of THz 2DCS to explore collective excitations inaccessible in other spectroscopies.
\end{abstract}

\maketitle
A transition to a state of matter that spontaneously breaks a continuous $U(1)$ symmetry of the Hamiltonian is characterized by the emergence of collective electronic modes associated with amplitude and phase fluctuations of the complex order parameter~\cite{Pekker2015,Shimano2020}.  The finite energy amplitude mode is an analog of the Higgs boson associated with the electroweak symmetry breaking of the fundamental vacuum~\cite{Pekker2015,Shimano2020}. 
In condensed matter physics, amplitude modes have been explored in various ordered phases, such as charge density waves (CDW)~\cite{Tsang1976,Demsar1999,Sugai2006,Yusupov2010}, antiferromagnets~\cite{Ruegg2008,Merchant2014,Jain2017}, and superfluid $^3$He~\cite{Lee1998,Volovik2014}. In the case of superconductors, long-range Coulomb interaction pushes the otherwise massless phase mode up to the plasma frequency, while the amplitude mode stays intact \textcolor{black}{at twice the SC gap energy \tdelta} \cite{Pekker2015,Shimano2020}. The observation of the amplitude mode in superconductors is challenging because it does not couple linearly to light being charge neutral, and it is expected to have a weak Raman response ~\cite{Cea2016R,Udina2022,BenfattoPRL2022} in cases when spontaneous Raman scattering can be interpreted as a probe of density fluctuations~\cite{Deveraux2007}. An exception is the case of 2$H$-NbSe$_2$~\cite{Sooryakumar1980,Littlewood1981,Littlewood1982,Measson2014,Grasset2018} or 2$H$-TaS$_2$~\cite{Grasset2019} where superconductivity coexists with CDW order and the amplitude mode can be seen in Raman spectroscopy by virtue of its coupling to the Raman-active soft CDW phonon~\cite{Cea2016,Grasset2018}.

\textcolor{black}{With the aim of identifying the amplitude mode in superconductors, the THz nonlinear optical response has been explored.} THz nonlinear responses, such as pump-probe response or third-harmonic generation (THG) in a conventional superconductor NbN, have been initially interpreted as the excitation of an amplitude mode through a nonlinear coupling to the electromagnetic field ~\cite{Matsunaga2014,Tsuji2015,matsunaga2015higgs,Matsunaga2017,Shimano2020}. Since then, the THz nonlinear response has been extensively explored in superconductors like MgB$_2$~\cite{Giorgianni2019,Kovalev2021,Reinhoffer2022}, high-temperature cuprates~\cite{Katsumi2018,Katsumi2020,Chu2020,Chu2023,Kim2023}, and iron-based systems~\cite{Vaswani2021,Isoyama2021,Grasset2022,Luo2022,Mootz2022}. Motivated by these experiments, theoretical works have shown that the third-harmonic (TH) THz nonlinear response is governed by quasiparticle excitations as well as the amplitude mode, with a relative hierarchy of the two contributions that depends on the disorder level and pairing strength ~\cite{Cea2016R,Jujo2018,Murotani2019,Silaev2019,Tsuji2020,Seibold2021,Fiore2022,Udina2022,Seibold2023,Puviani2023}. Importantly, theoretical considerations highlighted the possibility that THz nonlinearities can be mediated by an intrinsically different electronic response as compared to conventional nonlinearities at optical frequencies of a few eV ~\cite{Silaev2019,Udina2022,Seibold2021}. In principle, the latter is  governed by diamagnetic Raman-like density-density scattering in metals ~\cite{Deveraux2007}, whereas even weak disorder may allow THz light to trigger paramagnetic current-current fluctuations. Nevertheless, clear experimental indication of the predominance of paramagnetic processes has not been reported. 

Toward this aim, the recently developed multi-dimensional coherent spectroscopy is promising as it may allow one to disentangle different nonlinear processes~\cite{Cundiff2013}. In the case of THz two-dimensional coherent spectroscopy (2DCS), magnon~\cite{Lu2017}, phonon~\cite{Folpini2017}, plasmon~\cite{Houver2019}, and electronic excitations in disordered systems~\cite{Mahmood2021} have been clearly identified. In this Letter, we investigated the THz nonlinear response of the dirty-limit conventional superconductor NbN by THz 2DCS. Using broad-band THz, we observed a nonlinear signal peaked \textcolor{black}{at twice the SC gap energy \tdelta}. As the temperature increases, this nonlinear signal's peak exhibits a redshift following the temperature dependence of \tdelta obtained from the equilibrium optical response. To resolve the origin of these nonlinear spectral features, we employed narrow-band THz pulses at the driving angular frequency $\Omega/2\pi$~=~0.63~THz for THz 2DCS. We identified a first-harmonic (FH) nonlinear signal whose intensity displays a resonant enhancement when \tdelta matches the driving frequency $\Omega$. General theoretical considerations show that this FH signal, inaccessible in previous works focusing on THG, can only be generated by the aforementioned paramagnetic processes. Our numerical simulations demonstrate that with finite disorder, the $\Omega = 2 \Delta$ resonance of the FH response is dominated by the amplitude mode, offering a preferential route for its detection. 

\begin{figure}[t]
	\centering
	\includegraphics[width=\columnwidth]{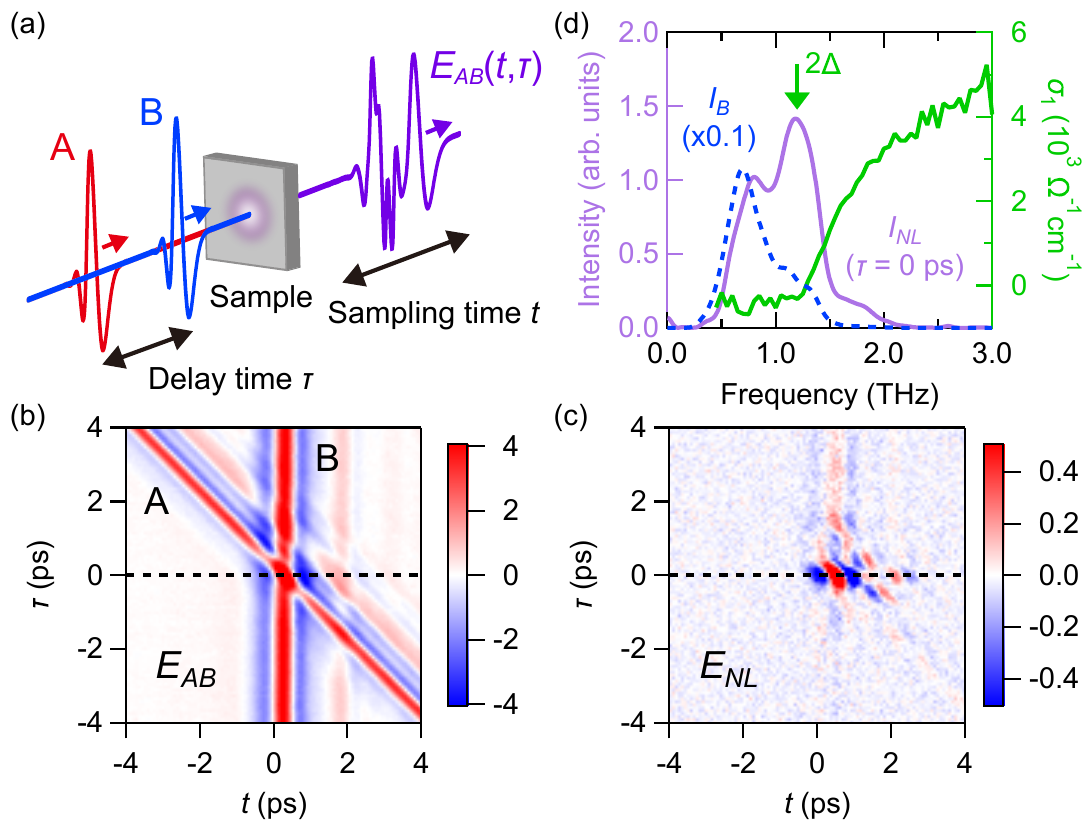}
	\caption{(a) A schematic of the THz 2DCS experiment. (b) Time traces of the A and B pulses together ($E_{AB}(t,\tau)$) transmitted after the NbN sample (\tc~=~14.9~K) at 5~K as a function of the sampling time $t$ and the delay time $\tau$. (c) The same plot as (b) but for the nonlinear signal $E_{NL}(t,\tau)$. (d) Power spectrum of the THz nonlinear signal from NbN at 5~K measured at $\tau=0$~ps with the broad-band THz pulses (purple, left axis). The blue dashed curve on the left axis is the power spectrum of the B-pulse. The green curve on the right axis shows the real part of the optical conductivity at 5~K.}
	\label{fig1}
\end{figure}

The schematic of the experiment is depicted in Fig.~1(a). Two intense single-cycle (broad-band) THz pulses are generated using the tilted-pulse front technique with LiNbO$_3$~\cite{Hebling2002,Watanabe2011,Hirori2011}. See the Supplemental Material (SM)~\cite{sm} for details. We first performed THz 2DCS on NbN (\tc~=~14.9 K) using broad-band THz pulses. The peak $E$-field of A and B pulses did not exceed 3~\kv to avoid the depletion of the SC condensate. We measured the transmitted THz $E$-fields of two pulses ($E_A(t,\tau)$, $E_B(t)$) separately, and of both pulses ($E_{AB}(t,\tau)$) together as shown in Fig.~1(b). We swept the delay time $\tau$ of the A-pulse with respect to the arrival of the B-pulse. The nonlinear signal's $E$-field is obtained as $E_{NL}(t,\tau) =E_{AB}(t,\tau) - E_{A}(t,\tau) - E_{B}(t)$, presented in Fig.~1(c). We perform a Fourier transform with respect to $t$, or both $t$ and $\tau$. Fig.~1(d) shows the power spectrum of the nonlinear signal at 5~K when $\tau=0$~ps (the purple curve). The nonlinear signal exhibits two peaks: one matches \textcolor{black}{twice the SC gap \tdelta} identified in the THz optical conductivity measured at 5~K (the green curve on the right axis), and the other is located at slightly higher energy than the center frequency of the B-pulse.

\begin{figure}[t]
	\centering
	\includegraphics[width=\columnwidth]{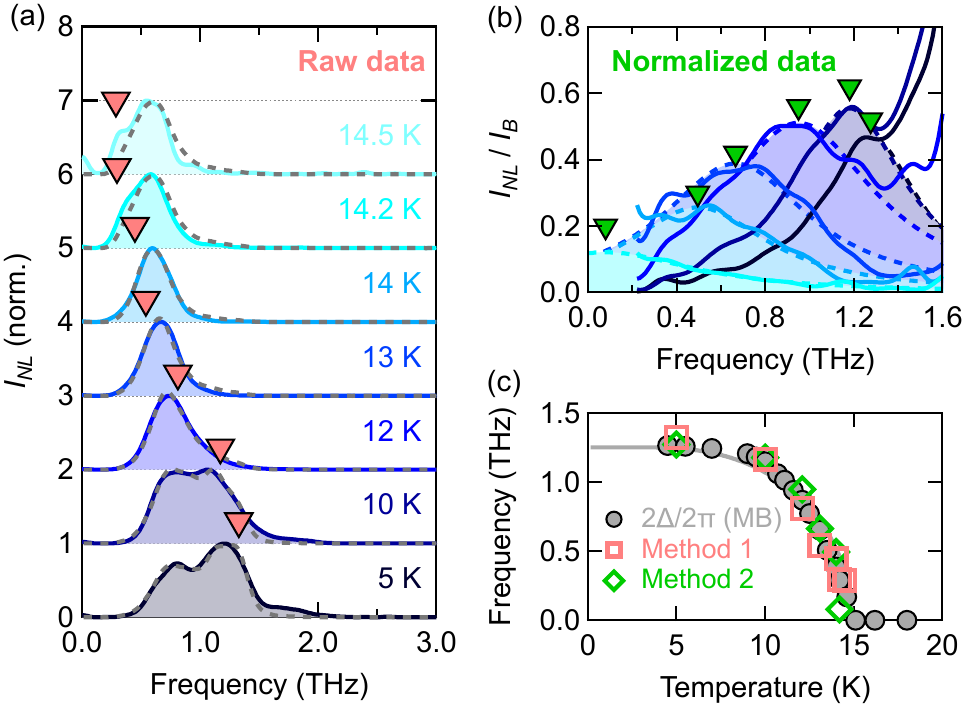}
	\caption{(a) Power spectrum of the nonlinear signal from NbN (\tc~=~14.9~K) when $\tau=0$~ps at different temperatures with single-cycle THz pulses. The gray dashed curves denote the fit to the data using a Lorentzian multiplied by the B-pulse power spectra. The obtained peak energy is presented by the red \textcolor{black}{arrows}. (b) The nonlinear signal in (a) divided by the B-pulse power spectra as a function of temperature. The obtained peak energy with fitting is shown by the green \textcolor{black}{arrows}. The dashed curves are the fits to the data using a Lorentzian. (c) Peak energy of the nonlinear signal as a function of temperature obtained by fitting the raw data in (a) (red open squares) and the normalized data in (b) (green open diamonds). The gray circles are the temperature dependence of the gap evaluated using the Mattis-Bardeen model.}
	\label{fig2}
\end{figure}

In Fig.~2(a), we present the power spectrum of the nonlinear signal at $\tau=0$~ps measured at different temperatures. The peak energy in the nonlinear signal shows a redshift with increasing temperature. We evaluated the energy of the peak in two ways. First, we fit the power spectra of the nonlinear signal using the power spectrum of the B-pulse multiplied by a Lorentzian. The fits reasonably reproduce the data as presented by the gray dashed curves in Fig.~2(a). The extracted peak energy is plotted as a function of temperature by red open squares in Fig.~2(c), in good agreement with the temperature dependence \textcolor{black}{of \tdelta} (the gray circles) as seen in the linear responses. We also evaluated the peak energy by normalizing the nonlinear signal spectra with the B-pulse spectra, as shown in Fig.~2(b). While the normalized spectra display an increasing tendency toward higher frequency below 12~K, likely due to the THG signal, peaks are discerned in the frequency range from 0.3 to 1.4~THz. We fit the normalized spectrum with a Lorentzian and plot the obtained peak energy as a function of temperature by green open diamonds in Fig.~2(c), consistent with those in the first procedure. \textcolor{black}{We note that the same result of the peak-energy shift is obtained using the A-pulse to normalize the nonlinear signal, as shown in Fig.~S2 in SM.} This direct correspondence between the peak energy of the nonlinear signal and the SC gap energy was also found in other NbN samples with different \tc, as presented in Fig.~S\textcolor{black}{7} in the SM.

\begin{figure}[t]
	\centering
	\includegraphics[width=\columnwidth]{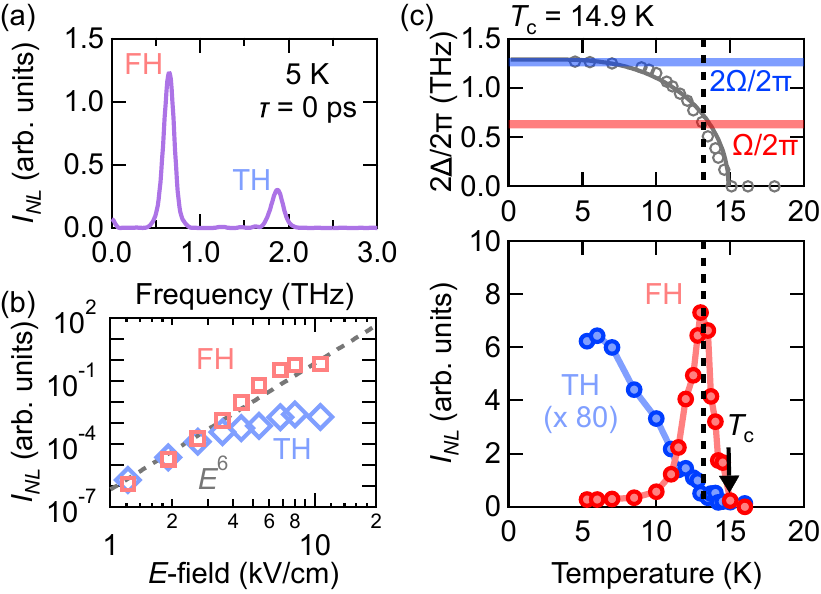}
	\caption{(a) Power spectrum of the THz nonlinear signal at 5~K measured at $\tau=0$~ps with narrow-band THz pulses. (b) The frequency-integrated intensity of the FH and TH contributions at 5~K when $\tau=0$~ps as a function of the B-pulse peak $E$-field. The dashed gray curve is the guide to the eye with a slope of 6. (c) Temperature dependence of the frequency-integrated intensity of the FH (red) and TH (blue) contributions. The top presents the temperature dependence of \tdelta evaluated from the linear conductivity as compared to $\Omega$ (red) and $2\Omega$ (blue). The solid gray curve is the gap computed by numerically solving the BCS equation. The black dashed line denotes the temperature that satisfies $\Omega=2\Delta$.}
	\label{fig3}
\end{figure}

To obtain deeper insight into the spectrum of the nonlinear signal, we performed THz 2DCS using narrow-band THz pulses with the peak $E$-field of 3~\kv and a driving angular frequency of $\Omega/2\pi$~=~0.63 THz. Figure~3(a) shows the nonlinear signal spectra for NbN with \tc~=~14.9~K measured at 5~K when $\tau=0$~ps. The power spectrum exhibits two peaks at 0.63~THz and 1.9~THz, corresponding to the FH and TH contributions, respectively. Both FH and TH intensities follow $E^6$ as shown in Fig.~3(b) up to the THz peak \textcolor{black}{$E$-field of} 10 and \textcolor{black}{3}~kV/cm, respectively, indicating that both are third-order nonlinear responses. In the third-order processes driven at a frequency $\Omega$, the nonlinear signals are generated at frequencies of $\pm\Omega\pm\Omega\pm\Omega$~\cite{Boyd}, giving signals at $\Omega$ and 3$\Omega$.


Figure~3(c) shows the frequency-integrated intensity of the FH and TH signals as a function of temperature with the multi-cycle THz pulses when $\tau=0$~ps. Here, we integrate from 0.3 to 1~THz for the FH signal, and from 1.6 to 2.2~THz for the TH signal. The temperature dependence of the TH signal is consistent with the previous reports where it takes its maximum intensity when {\it twice} the drive frequency matches \textcolor{black}{twice the SC gap energy \tdelta} (i.e. $2 \Omega = 2 \Delta$) as shown in the top panel~\cite{Matsunaga2014,Tsuji2015,Matsunaga2017,Shimano2020}. By contrast, the FH signal displays a resonant enhancement when the driving frequency matches \tdelta (i.e. $\Omega = 2 \Delta$). These behaviors were found in the other NbN samples with different \tc, as shown in Fig.~S\textcolor{black}{8} in SM. This FH signal has been overlooked in previous THz nonlinear experiments because it is usually overwhelmed by the transmitted $E$-field at the FH primarily from linear transmission, \textcolor{black}{as shown in SM}. The difference-signal analysis inherent to 2DCS allows it to be \textcolor{black}{unambiguously} isolated here. 

\begin{figure}[t]
	\centering
	\includegraphics[width=\columnwidth]{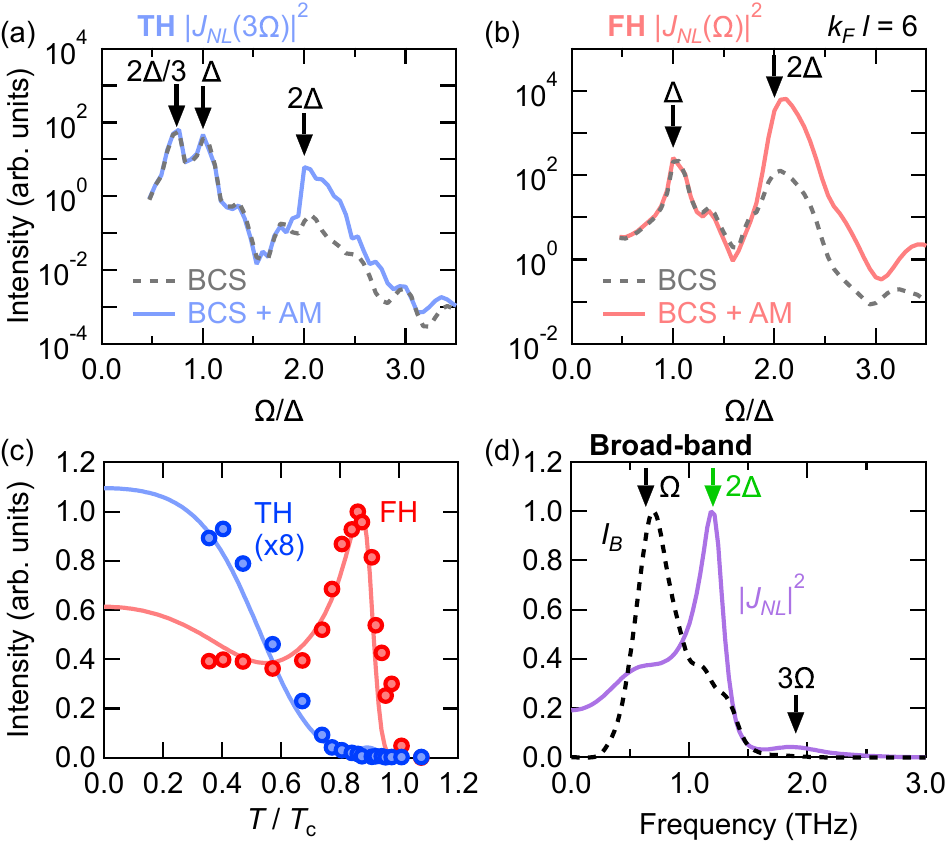}
	\caption{(a),(b) Nonlinear current intensity $|J_{NL}|^2$ for the (a) TH and (b) FH components as a function of the driving frequency $\Omega$ normalized by the SC gap $\Delta$ for the BCS contribution (dashed curve) and the sum of the BCS and amplitude-mode contributions (solid curves). (c) Comparison between the experimental data (open circles) and the theoretical results (solid curves) for the temperature dependence of the FH (red) and TH (blue). The temperature $T$ is normalized by \tc. To remove the temperature-dependent screening effects, the experimental nonlinear signal intensity ($|E_{NL}|^2$) is normalized by the internal B-pulse $E$-field ($E_B^6$) (see \cite{sm}). The theoretical results correspond to the nonlinear current intensity $|J_{NL}|^2$ at the driving frequency of 0.63~THz for a $E$-field constant in temperature.(d) The nonlinear current simulated using the broadband THz pulses. The black dashed curve shows the intensity of the driving THz $E$-field, obtained by simulating the experimental THz B-pulse. }
	\label{fig4}
\end{figure}

To clarify the experimental observations, we analyze the different nonlinear processes following the diagrammatic approach in previous work~\cite{Cea2016R,Silaev2019,Tsuji2020,Seibold2021}. The nonlinear processes in the SC state can be obtained by combining the diamagnetic (quadratic) and paramagnetic (linear) couplings between electronic excitation and the electromagnetic field. In the diagrammatic representation, the former is represented as a density-like electronic vertex with two-photon lines attached, while the latter appears as current-like electronic vertex with a single-photon line. The third-order nonlinear current is generally expressed as $J_{NL}(\omega) = \int \text{d}\omega_1 \text{d}\omega_2 \text{d}\omega_3 \delta(\omega_1+\omega_2+\omega_3-\omega) A(\omega_1)A(\omega_2)A(\omega_3) K(\omega_1,\omega_2,\omega_3) $, where $A(\omega)$ denotes the gauge field. Assuming monochromatic fields ($\omega_i=\pm\Omega$ for $i=1,2,3$), the possible combinations of $\omega_i$ gives peaks at  $\omega = \Omega$ and $\omega = 3\Omega$ in $J_{NL}(\omega)$. The spectral structure simplifies in the clean limit, when only diamagnetic processes are allowed \cite{Cea2016R}. {Because diamagnetic vertices have two-photon lines attached, the nonlinear kernel $K$} can be reduced to 
$K^{dia}(\omega_1+\omega_2)$ and its frequency permutations.  Given that the quasiparticles and amplitude-mode \textcolor{black}{fluctuations} are both enhanced at $2\Delta$, both the FH and TH diamagnetic responses are largest when $\omega_1+\omega_2 = 2 \Delta$ and thus for single-frequency driving at $\Omega$, their resonance occurs at $\Omega = \Delta$~\cite{Cea2016R,sm}. 

In contrast, when a finite amount of disorder is included, paramagnetic processes become allowed. Consequently, as detailed in the SM, the nonlinear kernel is a function $K^{para}(\omega_1,\omega_1+\omega_2,\omega_1+\omega_2+\omega_3)$, plus permutations of the running-frequencies, and one expects an enhancement whenever any of the three arguments of $K^{para}$ matches \tdelta. Therefore, unlike the resonance at $\Omega = \Delta$ for the FH and TH responses, the FH response at $\Omega = 2\Delta$, and TH responses at \textcolor{black}{$3\Omega = 2\Delta$ (i.e., $\Omega = 2\Delta/3$)} and $\Omega = 2\Delta$ can only arise via paramagnetic processes. \textcolor{black}{They are absent in the usual description of the spontaneous non-resonant Raman at optical frequencies, based on the widely-used effective-mass approximation which maps the Raman response into a diamagnetic-like susceptibility~\cite{Deveraux2007}. We note that the same paradigm does not necessarily apply to the resonant Raman case, leaving open e.g. the interpretation of recent non-equilibrium anti-Stokes Raman measurements in cuprate superconductors \cite{Glier2023}.}

We examined these expectations by numerical simulations using the attractive Hubbard model on a square lattice with Anderson-type impurities. Following the approach in Ref.~\cite{Seibold2021,Seibold2023}, we computed the nonlinear TH and FH signals, as shown in Figs. 4(a) and (b), respectively. Here, we set the disorder level to $k_Fl=$~\textcolor{black}{10} ($k_F$ is the Fermi wave vector and $l$ is the electronic mean free path) and evaluated the BCS-quasiparticle contribution (the dashed curves, denoted as BCS) and the sum of it and the amplitude-mode contribution (the solid curves, denoted as BCS + AM) separately (see SM for details). For the TH signal, the nonlinear current is enhanced at the expected three possible frequency combinations, as denoted by the vertical arrows in Fig.~4(a). {In the SM, we show that the relative ratio between the BCS and amplitude-mode contributions for the lower two peaks is consistent with previous reports~\cite{Silaev2019,Tsuji2020,Seibold2021}, where the amplitude mode acquires considerable spectral weight for very strong disorder.} On the other hand, as shown in Fig.~S\textcolor{black}{13}, we find that the upper peak at $\Omega=2\Delta$ is dominated by the amplitude mode for $k_Fl=3,6,10,$ and 30, which are relevant to the estimated disorder level of NbN \cite{Chand2012,Cheng2016} summarized in Fig.~S\textcolor{black}{12}. The FH $J_{NL}(\Omega)$ has two features that correspond to the two allowed frequencies for the FH response as discussed above.  \textcolor{black}{Again, the amplitude mode dominates the upper resonance for all disorder levels investigated, whereas the lower energy resonance shows appreciable disorder dependence.  We note that the dominance of the amplitude mode at $\Omega = 2\Delta$ in both TH and FH responses is an empirical observation from our numerics.  The reason for it is a topic of current investigation.}

Next, we compute the temperature dependence of the FH and TH signals at $\Omega/2\pi$~=~0.63~THz by modeling the kernel resonances numerically, using the temperature dependence of the SC gap obtained from the linear responses (see SM). Figure~4(c) compares the numerical results with the experimental data normalized by the incoming THz $E$-field for the FH and TH components. The calculated result for the FH component displays excellent agreement with experiments, unambiguously indicating that the FH signal stems from the amplitude-mode contribution at $\Omega = 2\Delta$ via the paramagnetic processes. The numerical result for the TH component displays a monotonic increase when the temperature is lowered, consistent with the experimental data. The increase in the TH intensity toward the lower temperature is due to the resonance of $\Omega = \Delta$ around 0~K, which is likely dominated by the amplitude mode in the high disorder level relevant for NbN. To check the internal consistency of our approach, we further simulate the nonlinear current driven by broad-band THz pulses (see SM for details). Figure~4(d) presents the nonlinear current induced by the broad-band THz pulses for $2\Delta/2\pi$~=~1.2~THz. The computed nonlinear current exhibits three peaks at $\Omega/2\pi = 0.63$~THz, $2\Delta/2\pi$~=~1.2~THz, and $3\Omega/2\pi = 1.9$~THz, qualitatively reproducing the experimental observation in Fig.~1(d).

Finally, to understand the full time evolution of the THz 2DCS response induced by single-cycle THz pulses for different delay times $\tau$, we modeled the time evolution of the nonlinear signal in close analogy with the previously discussed results (see SM). In particular, we modeled the largest contribution of the coherent response by a nonlinear kernel peaked at $2\Delta$, and additionally included a phenomenological incoherent kernel describing the out-of-equilibrium quasiparticle relaxation. Figure~S\textcolor{black}{19}(b) presents the simulated 2D spectra that qualitatively reproduces the data in Fig.~S\textcolor{black}{19}(a). \textcolor{black}{In the case of the narrow-band pulses, we can understand the THz 2D power spectra using the frequency-vector scheme \cite{Mahmood2021,Woerner2013}, as shown in SM. It is worth mentioning that the FH signals at $\omega_{\tau}=\pm\Omega$ are not resolvable in the strong-pump weak-probe experiments reported previously~\cite{matsunaga2015higgs}, as it derives from a single pump photon and two probe photons (see SM).}

In summary, we reported the THz nonlinear signal in conventional superconductors NbN using THz 2DCS. Using broad-band THz pulses, we identified a prominent spectral component of the nonlinear signal peaked \textcolor{black}{at twice the SC gap energy \tdelta}. With narrow-band THz pulses, the same nonlinear signal appeared at the driving THz frequency $\Omega$, in a fashion inaccessible in previous experiments. By varying the temperature, it manifests a resonant enhancement when $\Omega=2\Delta$. General theoretical analysis showed that this resonance at $\Omega=2\Delta$ in the nonlinear kernel can only arise from a paramagnetic coupling of photons to electronic current. Our numerical simulations revealed that this resonance in the paramagnetic nonlinear current is dominated by the amplitude mode of the SC order parameter.

Our work not only establishes the ability of THz 2DCS to access light-matter interactions unresolvable by previous THG or pump-probe experiments, but it also unambiguously demonstrates the intrinsic difference between the excitation processes contributing to the THz nonlinear response as compared to the optical nonlinearity that is usually understood by analogy to spontaneous non-resonant Raman spectroscopy. \textcolor{black}{These results open a novel perspective on the ability of THz 2DSC to detect and/or drive }other collective excitations such as the Leggett mode in multi-band superconductors~\cite{Giorgianni2019,Blumberg2007}, the Bardasis-Schrieffer mode in iron-based superconductors~\cite{Kretzschmar2013,Grasset2022}, or the Josephson plasmon in cuprate superconductors~\cite{Rajasekaran2016,Rajasekaran2018,Gabriele2021,NLwang2023,Kaj2023,Katsumi2023,Liu2023}.

We acknowledge Y. Gallais, S. Houver, A.-M. Tremblay, and R. Matsunaga for discussions. This project was supported by the Gordon and Betty Moore Foundation, EPiQS initiative, Grant No. GBMF-9454 and NSF-DMR 2226666, by the Sapienza University under Ateneo projects RM12117A4A7FD11B and RP1221816662A977. and by the European Union under grant ERC-SYN, MORE-TEM, GA 951215. K.K. was supported by Overseas Research Fellowship of the JSPS. N.P.A., L.B., and J.F. acknowledge the hospitality of the KITP (QUASIPART23 program), through support of NSF under Grant Nos. NSF PHY-1748958 and NSF PHY-2309135. P.R. and J.J. acknowledge the Department of Atomic Energy, Government of India (Grant 12-R \& D-TFR-5.10- 0100). G.S. acknowledges financial support from the Deutsche Forschungsgemeinschaft under SE 80/20-1.

\bibliography{NbN.bib}
\bibliographystyle{apsrev4-2}
\end{document}


\title{Revealing novel aspects of light-matter coupling by terahertz two-dimensional coherent spectroscopy: the case of the amplitude mode in superconductors
 Supplemental Material}
\author{Kota~Katsumi}
\email{kk5461@nyu.edu}
\affiliation{William H. Miller III , Department of Physics and Astronomy, The Johns Hopkins University, Baltimore, MD, 21218, USA} 
\author{Jacopo Fiore}
\affiliation{Department of Physics and ISC-CNR, Sapienza University of Rome, P.le A. Moro 5, 00185 Rome, Italy}
\author{Mattia Udina}
\affiliation{Department of Physics and ISC-CNR, Sapienza University of Rome, P.le A. Moro 5, 00185 Rome, Italy}
\author{Ralph~Romero III}
\affiliation{William H. Miller III , Department of Physics and Astronomy, The Johns Hopkins University, Baltimore, MD, 21218, USA} 
\author{David~Barbalas}
\affiliation{William H. Miller III , Department of Physics and Astronomy, The Johns Hopkins University, Baltimore, MD, 21218, USA} 
\author{John~Jesudasan}
\affiliation{Department of Condensed Matter and Material Science, Tata Institute of Fundamental Research, Homi Bhabha Rd., Colaba, Mumbai 400005, India}
\author{Pratap Raychaudhuri}
\affiliation{Department of Condensed Matter and Material Science, Tata Institute of Fundamental Research, Homi Bhabha Rd., Colaba, Mumbai 400005, India}
\author{Goetz Seibold}
\affiliation{Institut f\"{u}r Physik, BTU Cottbus-Senftenberg, P. O. Box 101344, 03013 Cottbus, Germany}
\author{Lara Benfatto}
\affiliation{Department of Physics and ISC-CNR, Sapienza University of Rome, P.le A. Moro 5, 00185 Rome, Italy}
\author{N.~P.~Armitage}
\affiliation{William H. Miller III , Department of Physics and Astronomy, The Johns Hopkins University, Baltimore, MD, 21218, USA} 
\maketitle
\section{Methods}
\subsection{Samples}
The NbN films were deposited on (100) oriented MgO substrates using reactive dc magnetron sputtering using a Nb target. The thickness of these films are approximately 20~nm. Disorder is tuned by controlling the level of Nb vacancies in the lattice, which is controlled by the Ar:N ratio in the sputtering chamber  \cite{Chockalingam2008}. The superconducting (SC) transition temperatures \tc of the samples were determined using two-coil mutual inductance measurements, as described in Ref. \cite{Chaudhuri2022}.

\subsection{Experimental details}
For terahertz (THz) two-dimensional coherent spectroscopy (2DCS), the output of a regenerative amplified Ti:sapphire laser with a center wavelength of 800~nm, pulse duration of 30~fs, pulse energy of 7~mJ, and repetition rate of 1~kHz, was divided into three beams: two for the generation of two intense THz pulses (A and B pulses), and the other for the gate pulse for the electro-optic (EO) sampling. The two THz pulses are combined to the same optical path using a silicon beam splitter. The transmitted THz pulses from the sample were detected by the EO sampling in a 1-mm-thick ZnTe (110) crystal up to 18~kV/cm \textcolor{black}{and a 0.5-mm-thick GaP  crystal for higher field strength to avoid the nonlinearity from the ZnTe}. We obtained peak THz electric-field ($E$-field) strengths of 72~\kv and 82~\kv for A and B THz single-cycle (broad-band) pulses.

\begin{figure}[t]
	\centering
	\includegraphics[width=\columnwidth]{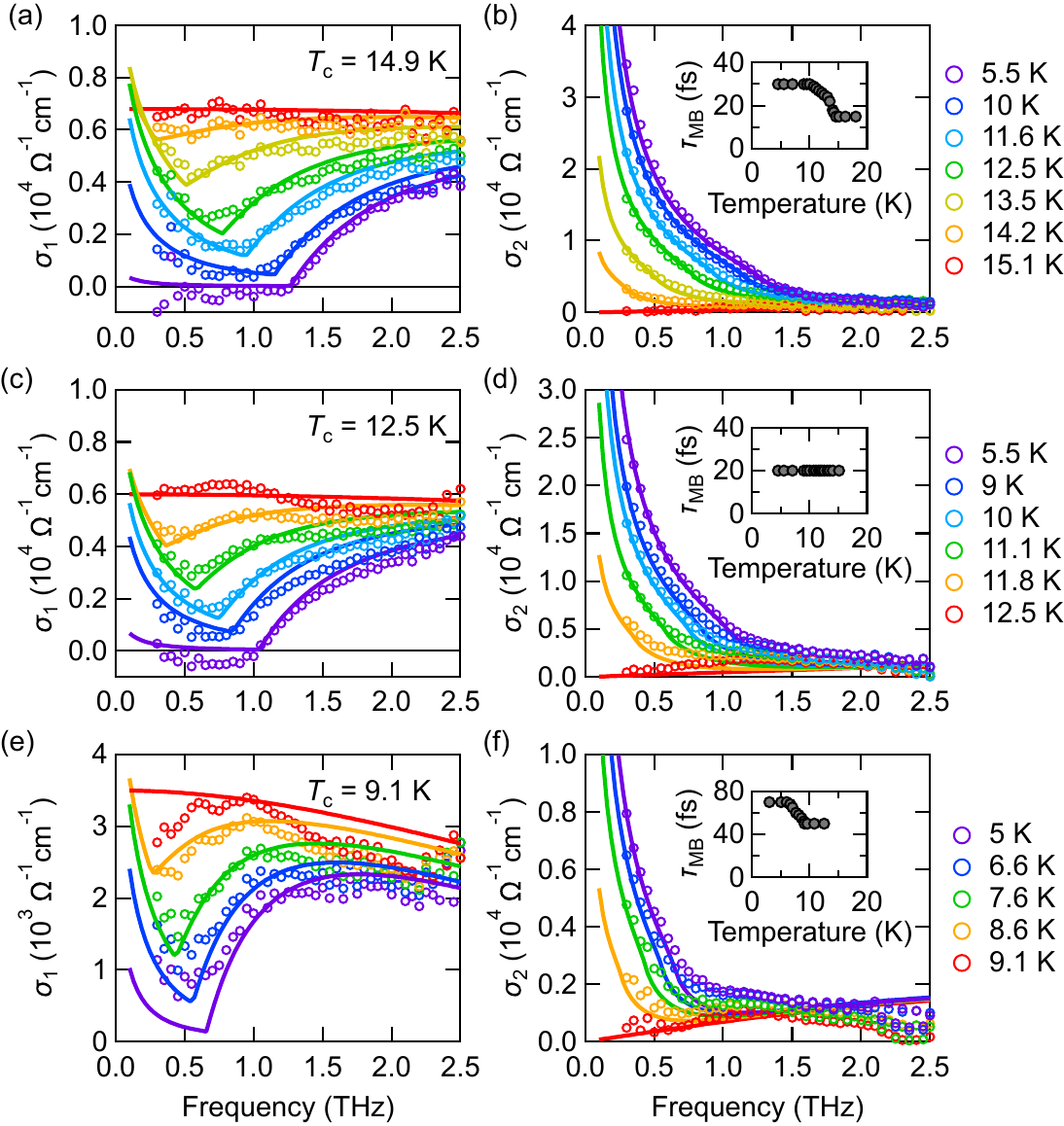}
	\caption{(a) Real and (b) imaginary part of the optical conductivity of NbN (\tc~=~14.9 K). The inset shows the fitted parameters of the relaxation time ($\tau_{\text{MB}}$) as Ref. \cite{Zimmermann1991}.  The solid curves are the optical conductivity computed using the Mattis-Bardeen \textcolor{black}{(MB)} model. (c), (d) The same plot as (a) and (b), respectively, but for NbN with \tc~=~12.5 K. (e), (f) The same plot as (a) and (b), respectively, but for NbN with \tc~=~9.1~K.}
	\label{figs1}
\end{figure}

The multi-cycle (narrow-band) THz pulses are generated using the band pass filters (BPFs) at the center frequency of $\Omega/2\pi$~=~0.63~THz, resulting in a peak THz $E$-field strength of 11~\kv for each pulse. \textcolor{black}{The choice of this frequency is because it straddles the ranges where the SC gap is found for a number of disorder levels. In this regard, instead of changing the driving frequency, we could choose the NbN samples with $T_{\text{c}}=14.9, 12.5,$ and 9.1~K whose SC gap energies \tdelta at 5~K are larger than $\Omega$ to identify the resonance of the FH signal at $\Omega=2\Delta$ when the temperature is increased.}

\section{Linear response and Determination of the superconducting gap energy}

To obtain the SC gap energy of the NbN films in equilibrium, we performed THz time-domain spectroscopy in the linear response regime. Figures \ref{figs1}(a) and (b) show the real and imaginary parts of the THz optical conductivity of the NbN sample with \tc~=~14.9~K. We evaluate the SC gap energy by the Mattis-Bardeen model \cite{Mattis1958,Zimmermann1991}. 
In Fig.~\ref{figs1}(a) and (b), the solid curves are the real and imaginary parts of the optical conductivity computed using the Mattis-Bardeen (MB) model. \textcolor{black}{The optical gap \tdelta}, which is twice the SC gap energy is obtained by the MB model and shown by circles as a function of temperature in the upper panel of Fig.~2(a) in the main text.  It follows \tdelta calculated by numerically solving the BCS gap equation (the solid curve). We performed the same analysis on even higher disorder NbN samples with \tc~=~12.5~K and \tc~=~9.1~K, whose results are shown in Figs. \ref{figs1} (c)-(d) and (e)-(f), respectively. The obtained SC gap energies of the NbN samples with \tc~=~12.5~K and \tc~=~9.1~K are plotted in the upper panels of Figs.~\ref{figs3}~(a) and (b), respectively.

\begin{figure}[t]
	\centering
	\includegraphics[width=\columnwidth]{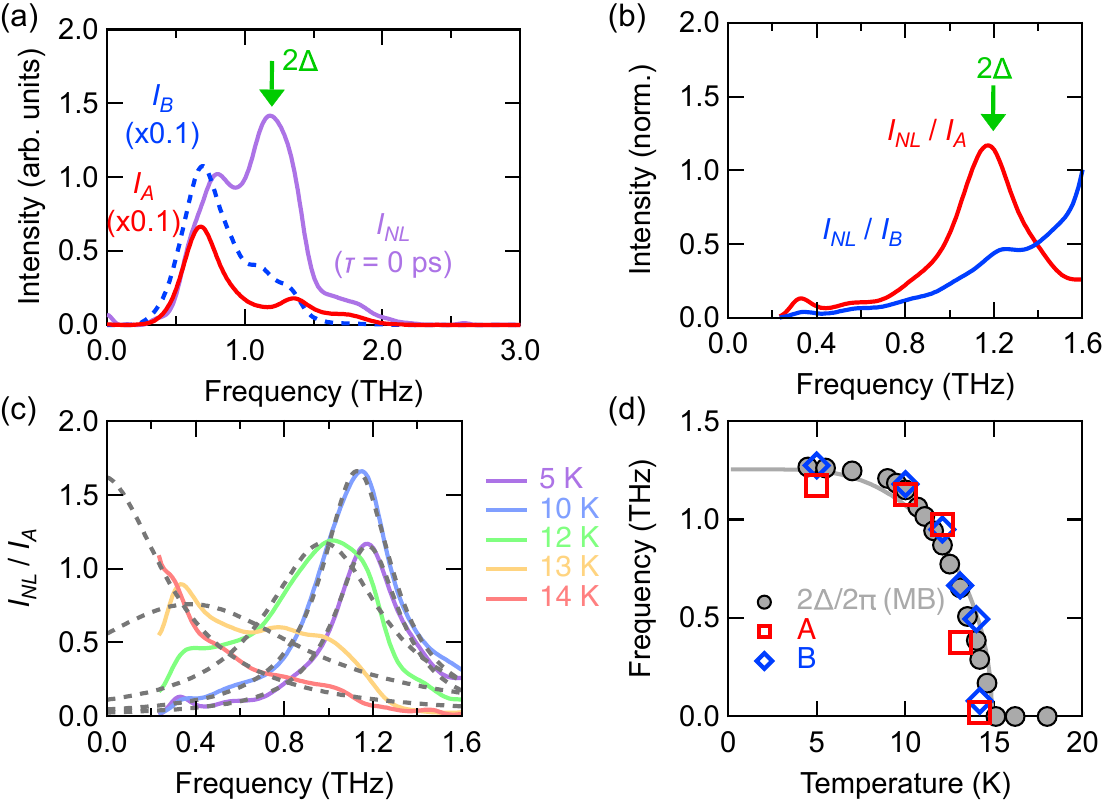}
	\caption{\textcolor{black}{(a) Power spectra of the A (red), B (blue), and nonlinear signal (purple) measured at 5 K. The green vertical arrow indicates \tdelta. (b) Power spectra of the nonlinear signal normalized by the A (red) and B (blue) pulse. (c) Temperature dependence of $I_{NL}/I_A$. The gray dashed curves are the Lorentzian fits to the data. (d) Peak energy of $I_{NL}/I_A$ (red) and $I_{NL}/I_B$ (blue) as a function of temperature. The gray circles are the temperature dependence of twice the SC gap evaluated using the MB model.}}
	\label{fig:norm_a}
\end{figure}

\textcolor{black}{
\section{Normalization of the nonlinear signal using the A pulse}}
\textcolor{black}{Since the A and B pulses are generated from the different LiNbO$_3$ crystals, their spectra are not exactly the same, as shown in Fig. \ref{fig:norm_a}(a), but they are very similar. In the main text, we chose B-pulse for normalizing the nonlinear signal in Fig.~2(b) as its spectrum is slightly smoother than that of the A-pulse. However, importantly, our analyzed results do not depend greatly on which pulse to use for normalization. Figure \ref{fig:norm_a}(b) compares the nonlinear signal $I_{NL}$ divided by $I_A$ (red) and $I_B$ (blue). While their relative magnitude and shape are slightly different, both peak positions match with \tdelta.  The temperature dependence of  $I_{NL}/I_A$ is shown in Fig. \ref{fig:norm_a}(c). The peak in $I_{NL}/I_A$ displays a red-shift when the temperature is increased. Figure \ref{fig:norm_a}(d) summarizes the peak energy obtained by fitting $I_{NL}/I_A$ with a Lorentzian as a function of temperature. The obtained peak energy from $I_{NL}/I_A$ shows a good agreement with that from $I_{NL}/I_B$. Therefore, this analysis and our conclusions do not depend on which pulse we normalize with.}

\textcolor{black}{
\section{Details of the fluence dependence of the nonlinear signal}}
\begin{figure}[t]
	\centering
	 \includegraphics[width=\columnwidth]{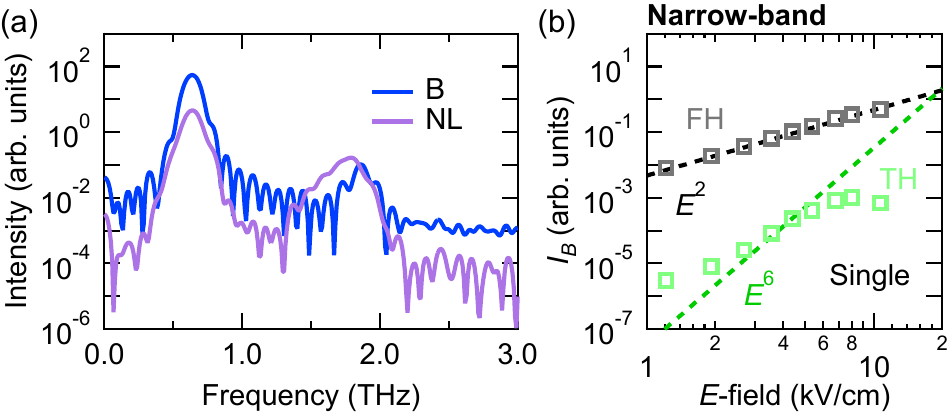}
	\caption{\textcolor{black}{(a) The power spectra of the narrow-band B-pulse (blue) and the nonlinear signal (NL, purple) from the NbN sample with \tc~=~14.9~K measured at 5~K. Here, the peak $E$-field of B-pulse is set to 4.4~kV/cm. (b) The $single$ B-pulse's THz $E$-field dependence of the FH (gray) and TH (light green) contributions in the transmitted B-pulse intensity $I_B$ measured at 5~K when the A-pulse is blocked. The dashed black and green lines are the guides to the eye with the slopes of 2 and 6, respectively.}}
	\label{fig:fluence_narrow}
\end{figure}

\textcolor{black}{We have seen that the nonlinear signal driven by the double THz pulses show the third-order nonlinearity, as presented in Fig.~3(b) in the main text. Here, we compare it with the case of the single-pulse experiment and the broad-band THz driving case. First, we discuss the case of the narrow-band THz driving for simplicity. Figure \ref{fig:fluence_narrow}(a) shows the power spectra of the B-pulse and the nonlinear signal from the NbN film with \tc~=~14.9~K measured at 5~K. The $single$ B pulse transmission shows the TH nonlinear signal, whose intensity roughly follows $E^6$ up to 4.4 kV/cm as plotted in Fig. \ref{fig:fluence_narrow}(b), consistent with the previous report by \cite{Matsunaga2014}. The deviation of the TH signal from $E^6$ dependence at higher $E$-field is due to the depletion of the SC condensate, and that below 2~kV/cm is most likely because of the leakage from the FH transmission, which follows $E^2$, as discussed in previous experiments, such as \cite{Katsumi2023}. However, the intensity of the FH signal is dominated by $E^2$ and the $E^6$ dependence is not discerned, as it is much smaller than the linear FH transmission, as evaluated in the following.}

\begin{figure}[t]
	\centering
	\includegraphics[width=5cm]{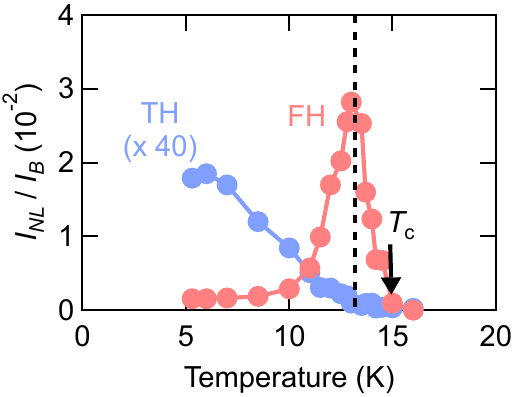}
	\caption{\textcolor{black}{Temperature dependence of the nonlinear signal intensity divided by that of the B-pulse for the FH (red) and TH (blue) components. The black vertical arrow indicates \tc, and the dashed black curve denotes 13~K, which satisfies $\Omega=2\Delta$.}}
	\label{fig:efficiency}
\end{figure}

\begin{figure}[t]
	\centering
	\includegraphics[width=\columnwidth]{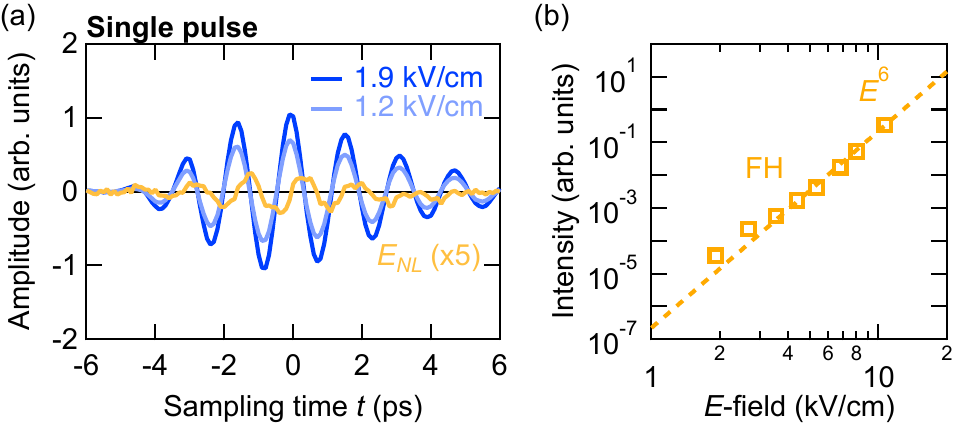}
	\caption{\textcolor{black}{(a) The single B-pulse’s wave forms for the incoming THz peak fields of $E$~=~1.9~kV/cm (blue) and $E_0$~=~1.2~kV/cm (right-blue) at 5~K with $E_{NL}=E_B(E)-E_B(E_0)E/E_0$ (orange). (b) Fluence dependence of the FH intensity in the power spectra of $E_{NL}$ at 5~K. The dashed line is a guide to the eye with the slope of 6.}}
	\label{fig:single_pulse}
\end{figure}

\textcolor{black}{Figure \ref{fig:efficiency} shows the nonlinear signal intensity $I_{NL}$ divided by that of the B-pulse $I_B$ for the FH (red) and TH (blue) components as a function of temperature using the data in Fig~ 3(c) in the main text. At this field the ratio $I_{NL}/I_{B}$ for the FH component takes its maximum value of $2.8 \times 10^{-2}$ at 13~K, when the resonant condition $\Omega=2\Delta$ is satisfied, whereas that for the TH component takes its maximum value of $4.5 \times 10^{-4}$ at 5~K. These maximum values indicate that the FH and TH nonlinear signals are 2-4 orders of magnitude smaller than the transmitted THz intensity.}

\textcolor{black}{In principle, we can perform $double$-pulse experiments if we have $two$ single-pulse data at THz $E$-fields of $E$ and $E_0$, and extract the nonlinear signal $E_{NL}$ by subtracting the linear transmission as $E_{NL}=E_B(E)-E_B(E_0)E/E_0$. Figure~\ref{fig:single_pulse}(a) presents the time trace of $E_{NL}$ for $E$~=~1.9~kV/cm and $E_0$~=~1.2~kV/cm. The FH intensity of $E_{NL}$ indeed follows $E^6$, as shown in Fig.~\ref{fig:single_pulse}(b). Importantly, while this procedure requires accurate calibration of $E/E_0$, THz 2DCS provides $E_{NL}$ in one measurement without controlling the THz $E$-field.}

\begin{figure}[t]
	\centering
	\includegraphics[width=\columnwidth]{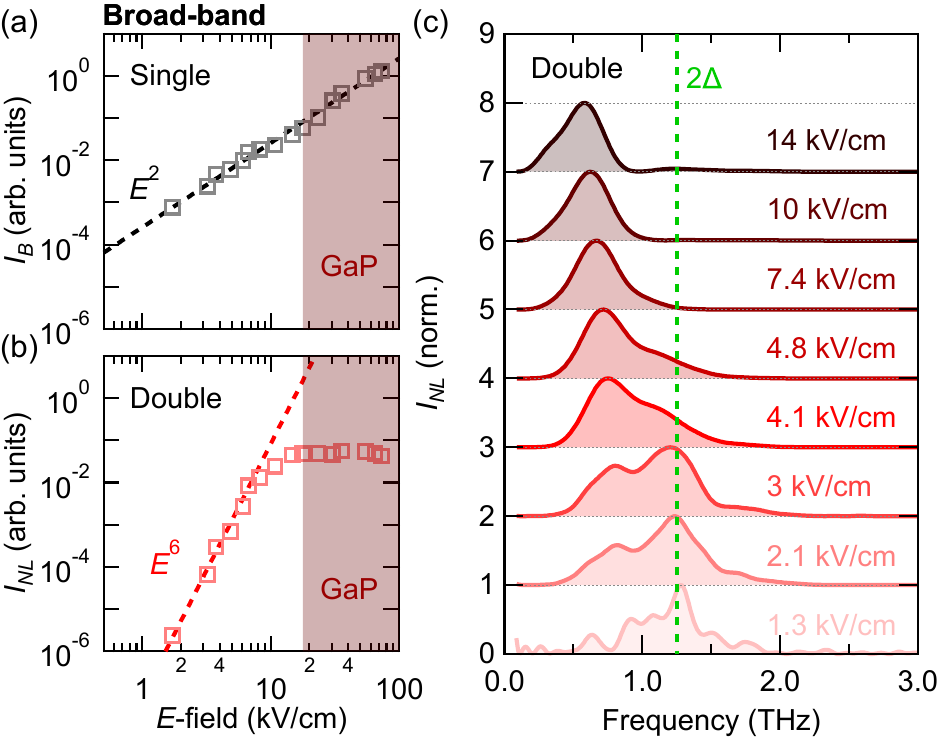}
	\caption{\textcolor{black}{(a) The transmitted broad-band $single$ B-pulse intensity $I_B$ as a function of the B-pulse's $E$-field measured at 5~K when the A-pulse is blocked. (b) The same plot as (a), but for the nonlinear signal when both A and B pulses are irradiated. (c) Fluence dependence of the nonlinear signal at 5~K. Each nonlinear signal is normalized by its maximum value for comparison. The green dashed vertical line indicates \tdelta in equilibrium.}}
	\label{fig:fluence_broad}
\end{figure}

One would expect a larger nonlinear signal for stronger THz $E$-field. To examine this hypothesis, we measured the fluence dependence of the transmitted B-pulse intensity up to 82~kV/cm when the A-pulse is absent, as presented in Fig.~\ref{fig:fluence_broad}(a). Here, we cannot disentangle the TH and FH components due to the broad-spectrum of the driving B-pulse. Again, the B-pulse transmission follows $E^2$ for all the $E$-field strength studied. On the other hand, the nonlinear signal using $double$ broad-band THz pulses, shown in Fig.~\ref{fig:fluence_broad}(b), displays the $E^6$ dependence up to 7~kV/cm, and completely saturates up to 82~kV/cm because of the depletion of the SC condensate.Therefore, the FH nonlinear signal cannot be obtained by the single-pulse experiment \textcolor{black}{with a single measurement} and is only accessible by \textcolor{black}{the two-pulse scheme}.

\textcolor{black}{Finally, we compare the power spectra of the nonlinear signal at 5 K as a function of the incoming THz $E$-field in Fig.~\ref{fig:fluence_broad}(c). With the $E$-field is increased, the peak in $I_{NL}$ displays a red-shift due to the reduction of \tdelta. This result further reinforces our assignment of the peak at \tdelta to the enhancement of the FH signal resonating at the amplitude-mode energy.}
\ 
\section{Nonlinear response with broad-band THz pulses for different samples}
\begin{figure}[t]
	\centering
	\includegraphics[width=\columnwidth]{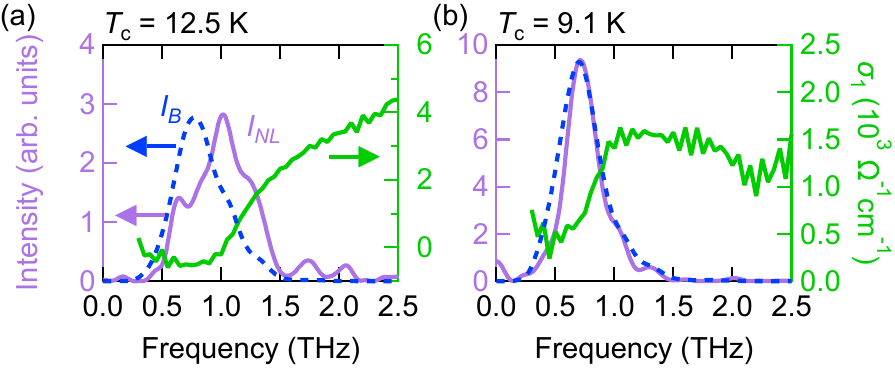}
	\caption{(a) Power spectrum of the THz nonlinear signal from NbN (\tc~=~12.5~K) at 5~K measured at the delay time $\tau=0$~ps with the broad-band THz pulses (purple, left axis). The blue dashed curve on the left axis shows the power spectrum of the B-pulse (multiplied by 0.002). The green curve on the right axis is the real part of the optical conductivity $\sigma_1(\omega)$ at 5~K. (b) The same plot as (a) but for the NbN with \tc~=~9.1~K. The power spectrum of the B-pulse (blue, left axis) is multiplied by 0.003.}
	\label{figs2}
\end{figure}

In addition to the experiments in the main text, we performed THz 2DCS measurements on NbN samples with \tc~=~12.5~K and 9.1~K using the broad-band THz pulses. Figure \ref{figs2} presents the power spectra of the nonlinear signal induced by the broad-band THz pulses (the purple curves) with the peak $E$-field of 1~\kv. The nonlinear signals at 5~K exhibit peaks at \textcolor{black}{twice the SC gap energy \tdelta} for both samples, consistent with the results for the sample with \tc~=~14.9~K in the main text.

\section{Nonlinear response with narrow-band THz pulse for different samples}
 In Fig.~\ref{figs3}, we show the temperature dependence of the nonlinear signal using the narrow-band THz pulses for the NbN samples with \tc~=~12.5~K and 9.1~K. While the TH component is not observed in the \tc~=~9.1~K sample, it is identified in the \tc~=~12.5~K sample and monotonically increases below \tc. We observe the FH component in both samples, showing resonant behavior when the driving frequency $\Omega$ matches \textcolor{black}{twice the SC gap energy \tdelta}. These behaviors are consistent with those in the \tc~=~14.9~K sample discussed in the main text.

\section{Normalizing the nonlinear signal by measured THz fields}
In this section, we describe how to normalize the nonlinear response to consider the fact that the $E$-field inside the sample depends on temperature. This largely arises from the temperature-dependent linear conductivity. Figure \ref{figs4} illustrates the multiple reflections of the THz $E$-field inside the NbN thin film on the MgO substrate in the presence of the incident THz $E$-field $E_{\text{incident}}(\omega)$. The THz $E$-field inside the sample $E_{\text{inside}}(\omega)$ at the position of $d_f/2$ is the sum of the $E$-field propagating to the left and right inside the film and can be written as
\begin{align}
	\label {eq1}
	E_{\text{inside}}(\omega)&=E_{\text{incident}}(\omega)\cfrac{2}{1+n_f(\omega)} \nonumber \\ &\cfrac{e^{in_f(\omega)d_f\omega/2c}+\cfrac{n_f(\omega)-n_s(\omega)}{n_f(\omega)+n_s(\omega)}e^{3in_f(\omega)d_f\omega/2c}}{1-\cfrac{n_f(\omega)-n_s(\omega)}{n_f(\omega)+n_s(\omega)}\cfrac{n_f(\omega)-1}{n_f(\omega)+1}\ e^{2in_f(\omega)d_f\omega/c}}.
\end{align}
Here, $c$ is the speed of the light, $d_f$ is the thickness of the film, and $n_f(\omega)$ and $n_s(\omega)$ are the complex refractive indexes of the thin film and substrate, respectively. If the thickness of the film is thin enough ($d_f \ll d_s, n_f(\omega)d_f\omega/c \ll 1$) and the refractive index of the film is much larger than that of the substrate ($n_f(\omega) \gg n_s(\omega)$), we can simplify Eq.~(\ref{eq1}) as
\begin{align}
	\label {eq2}
	E_{\text{inside}}(\omega)&=E_{\text{incident}}(\omega)\cfrac{2}{1+n_s(\omega)+Z_0\sigma(\omega)d_f},
\end{align}
where $Z_0$ is the impedance of free space, and we have used the relation between conductivity and refractive index for the film. We note that this expression is reminiscent of the usual transmission equation for THz light of a thin film on a substrate as \cite{Ikebe2009}
\begin{figure}[t]
	\centering
	\includegraphics[width=\columnwidth]{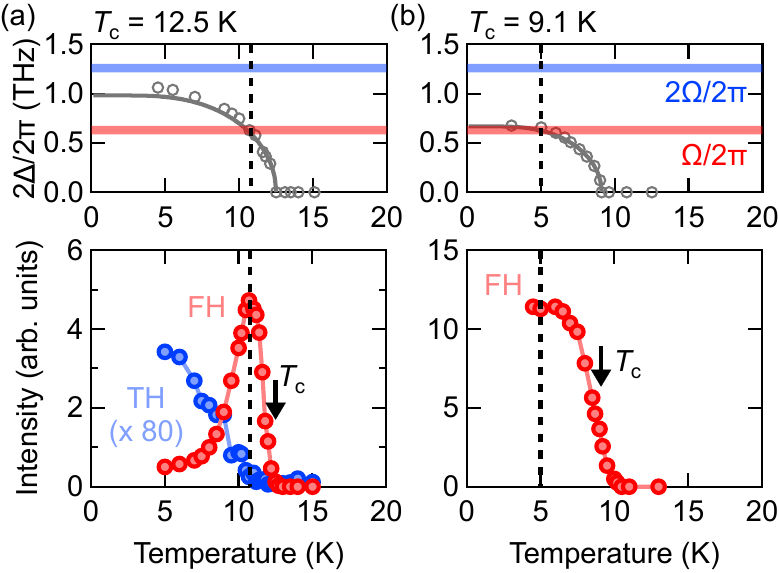}
	\caption{(a) Temperature dependence of the frequency-integrated intensity of the FH (red) and TH (blue) contributions, respectively for the NbN with \tc~=~12.5~K. The upper panel presents the temperature dependence of \tdelta evaluated from the equilibrium optical conductivity. The solid gray curve is \textcolor{black}{twice the SC gap energy} computed by numerically solving the BCS gap equation. The black dashed line shows the temperature that satisfies $\Omega=2\Delta$ (b) The same plot as (a) but for the NbN sample with \tc~=~9.1~K.}
	\label{figs3}
\end{figure}
Therefore, we can express the $E$-field inside the film simply in terms of the transmitted $E$-field as long as linear optical processes dominate. The $E$-field inside the film for the A-pulse is expressed by the transmitted $E$-field as
\begin{align}
	\label{eq4}
	E_{\text{inside}}^A(\omega)  &=  E^A_{\text{trans}}(\omega)
	\cfrac{n_s +1}{2n_s }     e^{-in_s d_s\omega/c},
\end{align}
where we have suppressed the frequency dependence of $n_s$ (and the same for the B-pulse). Therefore, we can reasonably approximate that the temperature dependence of $E^A_{\text{inside}}(\omega)$ is the same as that of $E^A_{\text{trans}}(\omega)$ because the substrate refractive index $n_s(\omega)$ does not depend strongly on temperature.

Given the information of $E^A_{\text{inside}}(\omega)$, one can derive an approximate normalization scheme for the nonlinear response. Nonlinear processes drive a time-dependent nonlinear sheet current in the film that can be expressed as
\begin{align}
	\label{Sheet}
	\kappa^{(3)}(\omega) =  G^{(3)} E^A_{\text{inside}}(\omega_1) E^B_{\text{inside}}(\omega_2) E^C_{\text{inside}}(\omega_3),
\end{align}
where $ G^{(3)}$ is a nonlinear conductance, and the electric fields corresponding to the three driving fields inside the film are given by Eq. (\ref{eq4}). The driven nonlinear current is unambiguous if one has independent control of the three frequencies. In the current experimental situation of the narrow-band THz pulses peaked at $\Omega$, the resultant signal will be the full mixture of any frequencies such that $\omega = \pm \Omega \pm \Omega   \pm \Omega$. The $E$-field emitted into the substrate from a (nonlinear) sheet current at frequency $\Omega$, i.e. the FH component is given by
\begin{align}
	\label{antenn1}
	E_{NL}(\Omega ) =  \kappa^{(3)} (\Omega ) \frac{ Z_0}{1+n_s} .
\end{align}
THz light inside the substrate is transmitted to vacuum with an additional factor of $2 n_s  e^{in_s d_s\Omega/c} / (1 + n_s) $ giving a transmitted field out of the substrate as
\begin{align}
	\label{antenn2}
	E_{NL}(\Omega ) =  \kappa^{(3)} (\Omega ) \frac{ 2 n_s Z_0 e^{in_s d_s\Omega/c}}{(1+n_s)^2}.
\end{align}
Combining Eqs. (\ref{eq4}), (\ref{Sheet}), and (\ref{antenn2}), we obtain 
\begin{align}
	\label{final}
		G^{(3)} = \frac{1}{D} \frac{E_{NL}(\Omega) }{E_{\text{trans}}^A E_{\text{trans}}^A E_{\text{trans}}^B} \frac{(2n_s)^2}{1+n_s} \frac{e^{2i n_s d_s\Omega/c}}{ Z_0},
\end{align}
where again we suppressed the weak frequency dependence of $n_s$. Here, $D=3$ is the degeneracy factor that accounts for the number of unique permutations that two pulses $A$ and $B$ can contribute to Eq. \ref{Sheet}. \textcolor{black}{The TH component of the $E$-field emitted from the nonlinear sheet current can also be expressed by Eq. \eqref{final} but with the degeneracy factor of $D=1$.}

\begin{figure}[t]
	\centering
	\includegraphics[width=8cm]{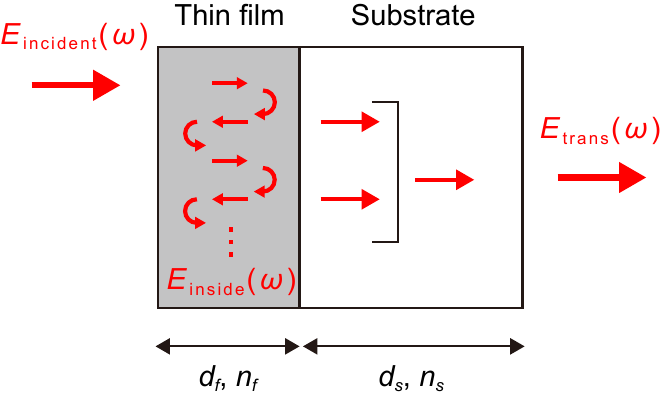}
	\caption{A schematic illustration of the transmitted THz $E$-field for a thin film on a substrate.}
	\label{figs4}
\end{figure}
\begin{align}
	\label {eq3}
	E_{\text{trans}}(\omega)&=E_{\text{incident}}(\omega)\cfrac{4n_s(\omega)}{1+ n_s(\omega)} \; \cfrac{e^{in_s(\omega)d_s\omega/c}}{1+n_s(\omega)+Z_0\sigma(\omega)d_f}.
\end{align}

\begin{figure*}[t]
	\centering
	\includegraphics[width=15cm]{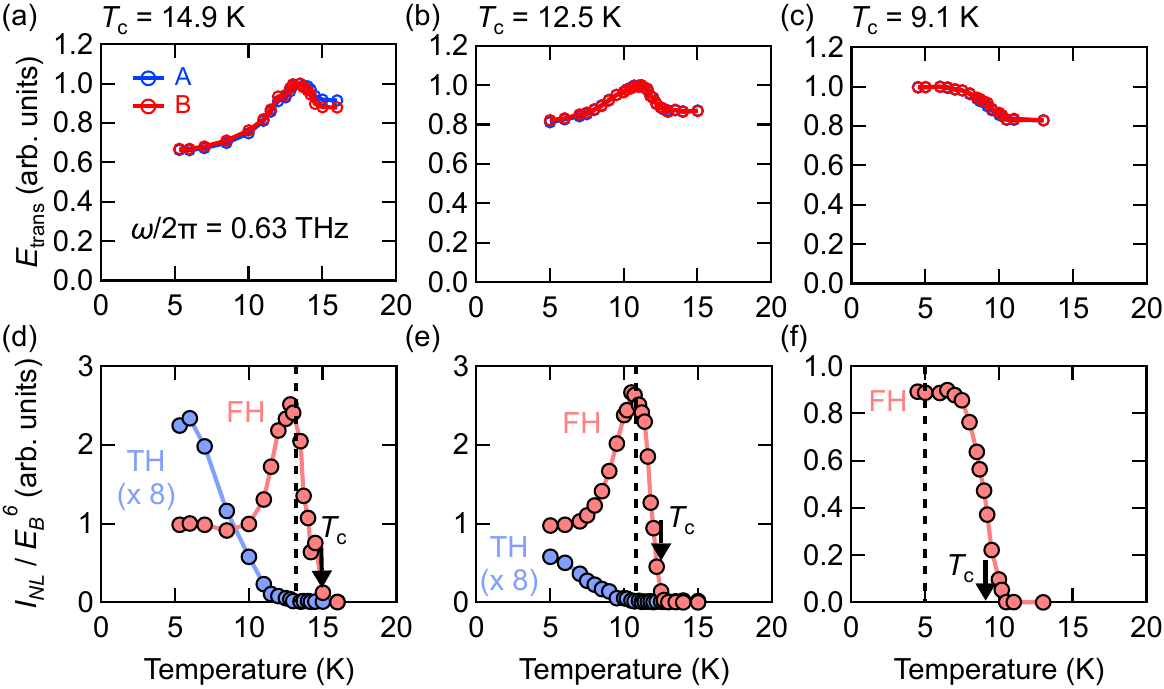}
	\caption{(a) Temperature dependence of the transmitted $E$-field from the NbN sample with \tc~=~14.9~K at 0.63~THz. The transmitted  spectra is integrated from 0.3 to 0.9~THz. (b), (c) The same plot as (a) but for the NbN samples with \tc~=~12.5~K and 9.1~K, respectively. (d) Frequency-integrated intensity of the FH (red) and TH (blue) signals for the NbN sample with \tc~=~14.9~K normalized by $E_{\text{trans}}(\Omega)^6$ of the B-pulse. \textcolor{black}{Note that Fig.~\ref{fig:efficiency} is normalized by $I_{\text{trans}}(\Omega) = E_{\text{trans}}(\Omega)^2$.} The vertical black dashed line denotes the temperature satisfying $\Omega=2\Delta$. The vertical black arrow indicates the \tc of the sample. (e), (f) The same plot as (d) but for the NbN samples with \tc~=~12.5~K and 9.1~K, respectively.}
	\label{figs5}
\end{figure*}

In Figs.~\ref{figs5}(a)-(c), we present the temperature dependencies of $E^A(\Omega)$ and $E^B(\Omega)$ for the NbN samples with \tc~=~14.9~K, 12.5~K, and 9.1~K. respectively.  The temperature dependence is seen to small overall.   Moreover, since the temperature dependence of $E^A$ matches $E^B$ in all the samples studied, we can normalize the nonlinear intensities using $E^B$ alone as $[E^B(\Omega)]^6$. Such a normalization assumes that only degenerate frequency mixing is occurring. Figures~\ref{figs5}(d)-(f) summarize the normalized nonlinear signals as a function of temperature.  The corrected nonlinear signals also exhibit a resonant enhancement when $\Omega=2\Delta$, consistent with that seen in the raw data shown in Fig.~3(c) in the main text and Fig.~\ref{figs3}.

\section{Details of the theoretical analysis}

\subsection{Formalism}
The third-order nonlinear current proportional to the gauge field can be written in the most general form as \footnote{In order to avoid lengthy expressions, we note that whenever in the text we write a kernel which is not symmetric with respect to the exchange of the indices $1,2,3$, the expression in Eq.\ (\ref{eq:genker}) gets modified as $K(\omega_1,\omega_2,\omega_3)=1/3!\sum_{\pi\in S_3}K(\omega_{\pi(1)},\omega_{\pi(2)},\omega_{\pi(3)})$, where $\pi$ is an element in the symmetric group $S_3$ of the possible permutations of three indices.}
\begin{align}
\label{eq:genker}
J_{NL}(\omega)&=\int \text{d}\omega_1 \text{d}\omega_2 \text{d}\omega_3\, \delta(\omega-\omega_1-\omega_2-\omega_3) \nonumber\\
&K(\omega_1,\omega_2,\omega_3) A(\omega_1)A(\omega_2)A(\omega_3).
\end{align}
For an incoming monochromatic field $A(\omega)=A_0\left[\delta(\omega-\Omega)+\delta(\omega+\Omega)\right]$, the nonlinear current in Eq.~(\ref{eq:genker}) leads to
\begin{align}
J_{NL}(\omega)&=\delta(\omega-\Omega)[K(\Omega,\Omega,-\Omega)+K(\Omega,-\Omega,\Omega) \nonumber\\
&+K(-\Omega,\Omega,\Omega)]A_0^3 \nonumber\\
&+\delta(\omega-3\Omega)K(\Omega,\Omega,\Omega)A_0^3+ (\Omega\to-\Omega).
\end{align}
This predicts the generation of two harmonics at frequencies $\Omega$ and $3\Omega$, both of which can be accessed by THz 2DCS as described herein.

\begin{figure}[h]
\includegraphics[width=\columnwidth]{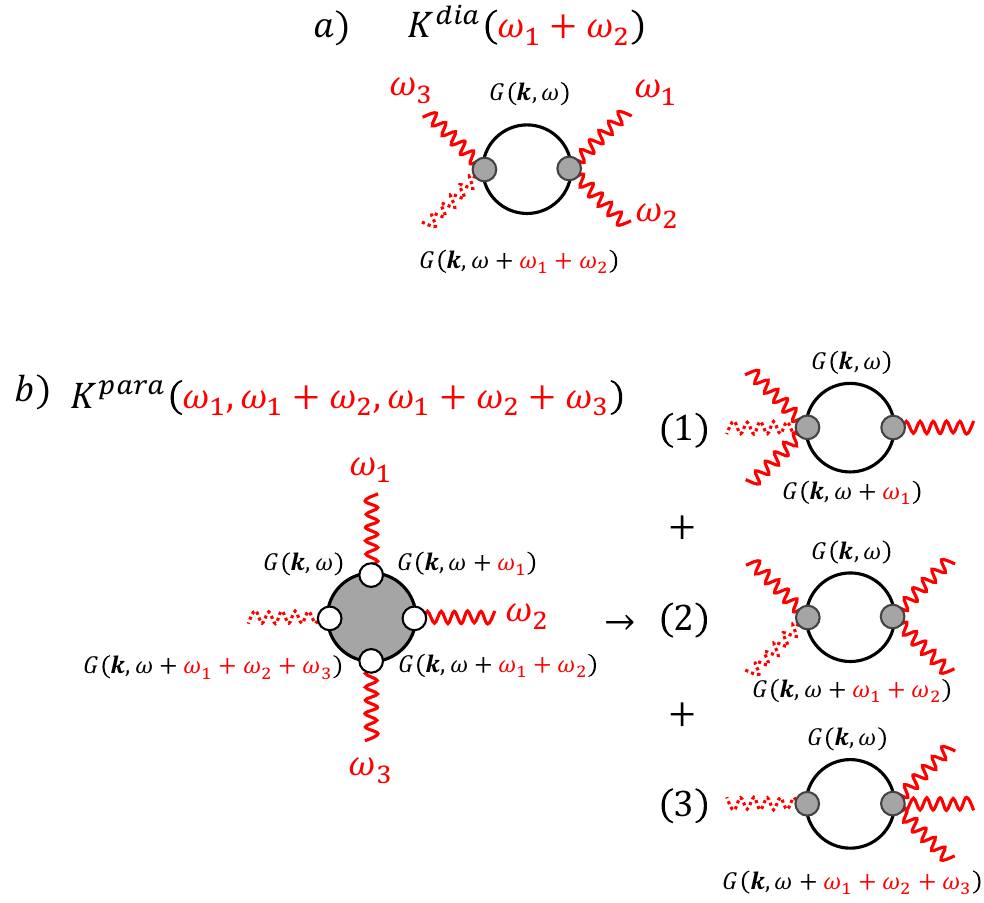}
\caption{Diagrams that contribute to the purely BCS part of the nonlinear kernel. In the clean case \textcolor{black}{(a)}, $K^{dia}$ is the only important process, involving two diamagnetic interactions (grey bullets). This produces a response that explicitly depends only on the sum of two incoming photons (red wavy lines) entering a convolution of two fermionic propagators (black lines) $G(\mathbf{k},\omega)$ \textcolor{black}{where $\omega$ and $\mathbf{k}$ are the frequency and momentum, respectively, of the electronic Green's function, over which one integrates.} In the disordered case \textcolor{black}{(b)}, the paramagnetic process $K^{para}$ dominates the nonlinear response, with four current-like vertices (white bullets) connecting four electron propagators and four-photon lines. In this case, one has different combinations of the external frequencies entering the fermionic propagators, which are convoluted. The grey-shaded area inside the bubble represents the self-energy and vertex corrections induced by impurities. \textcolor{black}{The full frequency structure of $K^{para}$ on the left is mapped through Eq.~(\ref{eq:kt}) into the sum of the three effective Kubo-like susceptibilities on the right.}}
\label{fig:paradia}
\end{figure}

As widely established~\cite{Cea2016}, the general expression Eq.\ (\ref{eq:genker}) for a clean superconductor is greatly simplified because only a diamagnetic, Raman-like density interaction between fermions and gauge fields is allowed.  The relevant diagram is shown in Fig.~\ \ref{fig:paradia} \textcolor{black}{(a)}. Accordingly, one has only two-photon vertices and Eq.~(\ref{eq:genker}) becomes
\begin{align}
J_{NL}^{dia}(\omega)&=\int \text{d}\omega_1 \text{d}\omega_2 \text{d}\omega_3\, \delta(\omega-\omega_1-\omega_2-\omega_3) \nonumber\\
&K^{dia}(\omega_1+\omega_2)A(\omega_1)A(\omega_2)A(\omega_3)\\
&=\int d\omega^{\prime}\,K^{dia}(\omega^{\prime})A(\omega-\omega^{\prime})A^{(2)}(\omega^{\prime}),
\end{align}
having denoted $A^{(2)}(\omega)\equiv \int \text{d}\omega^{\prime}\,A(\omega^{\prime})A(\omega-\omega^{\prime})$. Here, the only possible maxima in the nonlinear current occurs when the sum of \textit{two} incoming photon frequencies match the energy at which the kernel is enhanced, i.e., for a superconductor $\omega_1+\omega_2=2\Delta$. In the case of monochromatic incident pulses at frequency $\Omega$, this condition is satisfied when $2\Omega=2\Delta$, at which both the FH and TH responses are expected to be enhanced. By contrast, in our experiments, the nonlinear FH response in the disordered NbN is enhanced at $\Omega=2\Delta$, indicating that the diamagnetic process cannot account for our observation, leading us to consider other processes.

\begin{figure}
\includegraphics[width=\columnwidth]{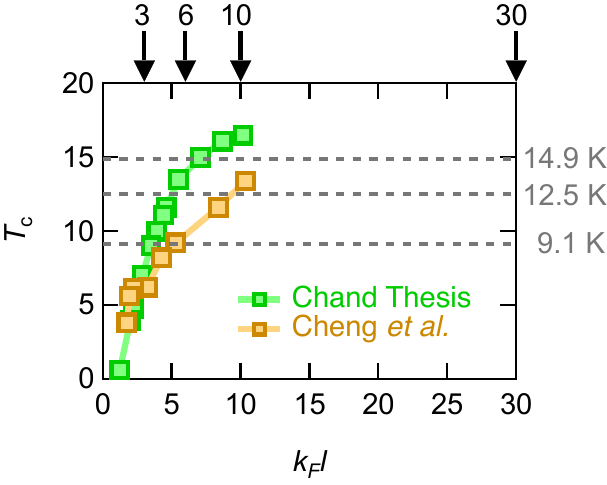}
\caption{\textcolor{black}{The SC transition temperature \tc of NbN samples versus the disorder level $k_Fl$ in Refs. \cite{Chand2012,Cheng2016}. The dashed gray horizontal lines indicate the \tc of the NbN samples used in our study. The black horizontal arrows denote the $k_Fl$ values used in our simulations.
}}
\label{fig:kfl}
\end{figure}

\begin{figure*}
\includegraphics[width=\linewidth]{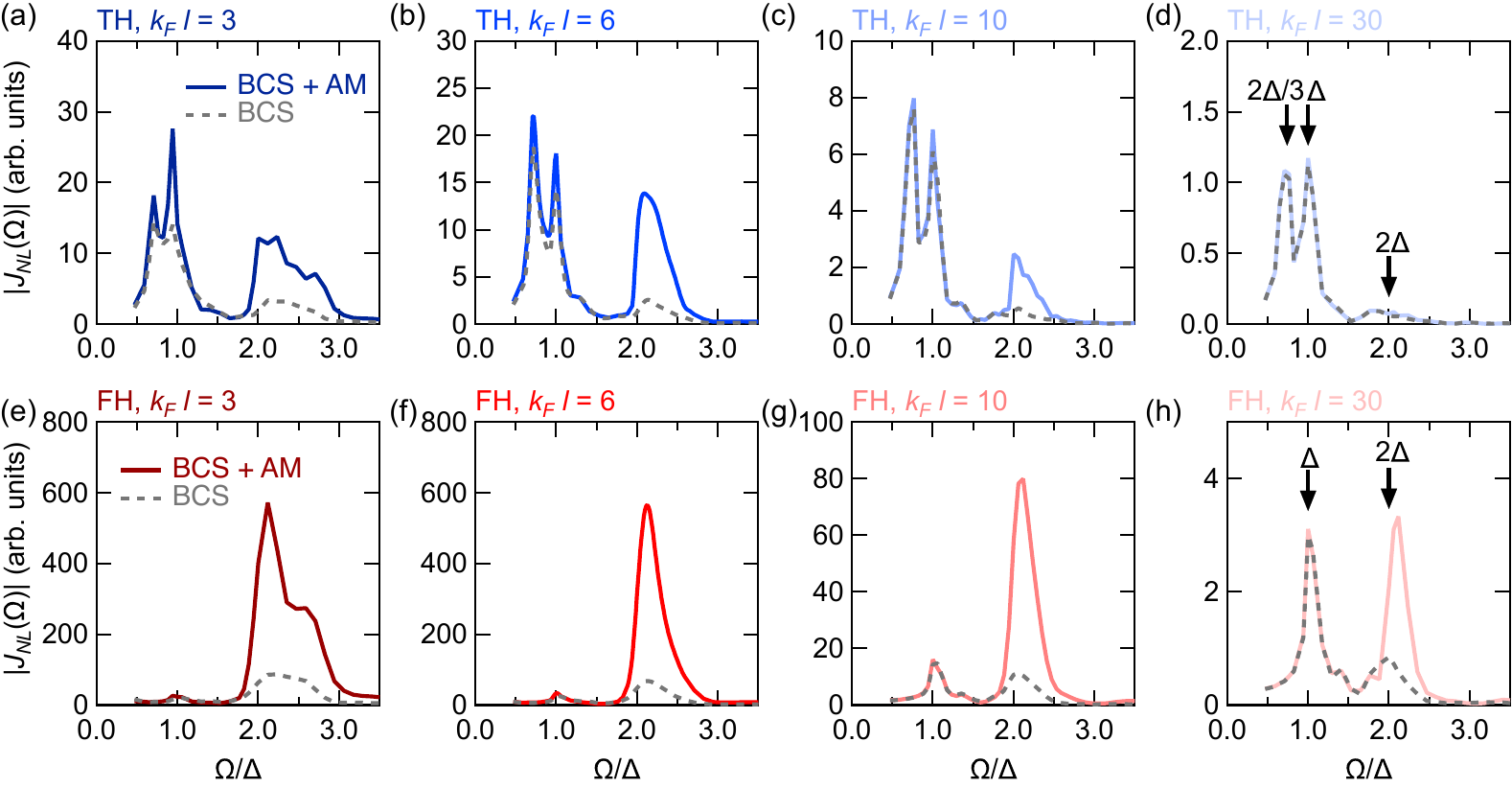}
\caption{\textcolor{black}{(a-d) The nonlinear current amplitude $|J_{NL}(\Omega)|$ for the TH components as a function of the driving frequency $\Omega$ normalized by the SC gap $\Delta$, for $k_Fl$ values of (a) 3, (b) 6, (c) 10, and (d) 30, respectively. The BCS contribution and the sum of the BCS and amplitude-mode (AM) contributions are denoted by the dashed and solid curves, respectively. (e-h) The same plots as (a-d), but for the FH components.}}
\label{fig:sm}
\end{figure*}

In the presence of disorder, coupling via paramagnetic current-like processes are also allowed, which includes both self-energy and vertex corrections induced by scattering with impurities. In previous theoretical works \cite{Seibold2021,Benfatto2023,Tsuji2020,Silaev2019}, it has been shown that the most important contribution to the nonlinear current comes from the diagram on the \textcolor{black}{left} in Fig.~\ref{fig:paradia} \textcolor{black}{(b)}, where one has to retain the full generic expression in Eq.\ (\ref{eq:genker}), plus vertex corrections induced by amplitude and phase fluctuations, which dominate the signal at high disorder levels. By examining the structure of this diagram, we can rewrite the nonlinear kernel in the form of $K^{para}(\omega_1,\omega_1+\omega_2,\omega_1+\omega_2+\omega_3)$, leading to
\begin{align}
\label{eq:jpara}
J_{NL}^{para}(\omega)=\int \text{d}\omega_1 \text{d}\omega_2 \text{d}\omega_3\, \delta(\omega-\omega_1-\omega_2-\omega_3) \nonumber\\
K^{para}(\omega_1,\omega_1+\omega_2,\omega_1+\omega_2+\omega_3)A(\omega_1)A(\omega_2)A(\omega_3).
\end{align}
\textcolor{black}{From Eq.~\eqref{eq:jpara}, one can find that for a monochromatic field with frequency $\Omega$, the paramagnetic nonlinear current displays resonances whenever any of the three combinations of external frequencies running into the fermionic kernel in $K^{para}(\omega_1,\omega_1+\omega_2,\omega_1+\omega_2+\omega_3)$ match $2\Delta$. Table \ref{tab1} summarizes the resonance conditions of the TH and FH paramagnetic nonlinear responses for the monochromatic incoming fields at $\Omega$. Importantly, the resonances at \textcolor{black}{($3\Omega=2\Delta$ i.e., $\Omega=2\Delta/3$)} and $2\Delta$ in the TH component and at $\Omega=2\Delta$ in the FH component are unique to the paramagnetic process. In particular, the resonance of the FH paramagnetic contribution at $\Omega=2\Delta$ coincides with our experimental observation, unambiguously demonstrating that the observed FH nonlinear signal stems from  paramagnetic processes.}

\subsection{Numerical simulations}
\begin{table}[t]
\caption{Resonance conditions for the TH and FH nonlinear kernels in the diamagnetic and paramagnetic processes.}
    \centering
    \begin{tabular}{ |c| c| c| }
        \hline
        &TH&FH \\
        \hline
        \ $K^{dia}$ \ & \ $\Omega=\Delta$ & \  $\Omega=\Delta$ \ \\ 
        \hline
        \ $K^{para}$ \ & \ $\Omega=2\Delta/3,\Delta,2\Delta$ \ & \ $\Omega=\Delta,2\Delta$ \  \\ 
        \hline
    \end{tabular}
    \label{tab1}
\end{table}
Having established the theoretical framework, we simulated the time evolution of an attractive Hubbard model with local disorder\textcolor{black}{
\begin{equation}\label{eq:ahub}
H=-t\sum_{\langle ij\rangle} c^\dagger_{i,\sigma}c_{j,\sigma}
-|U|\sum_i n_{i,\uparrow}n_{i,\downarrow}+\sum_{i,\sigma} V_i n_{i,\sigma}\,.
\end{equation}
Here, the first term describes the hopping of charge carriers
between nearest neighbors, the second term is the on-site attractive interaction, and the last term introduces disorder via a site-dependent potential.
The latter is taken from a flat distribution with $-V_0 \le V_i \le +V_0$.
In addition, we coupled a monochromatic gauge field to Eq.~(\ref{eq:ahub}) via the Peierls substitution and solved the model by evaluating the time-dependent Bogoliubov-De Gennes equations. Nonlinear FH and TH components of the nonlinear current were then computed from the resulting time-dependent density matrix. Results presented here were obtained for a $16\times 16$ square lattice at zero temperature with charge carrier concentration and attractive interaction fixed to $n=0.5$ and $U/t=2$, respectively. The strength of the impurity potential $V_0/t$ is converted to the dimensionless parameter $k_Fl$ where $k_F$ is the Fermi wave vector and $l$ is the electronic mean free path.}  The details of the numerical implementation are described in Refs.\ \cite{Seibold2021, Benfatto2023, Seibold2023}. In the numerical simulations, we can separately evaluate the nonlinear current, which arises from only BCS quasiparticle excitations and that which also includes the amplitude mode, by constraining the time-dependence of the SC order parameter in the density matrix or not. In order to confirm the dependence of the results on the strength of the disorder, we performed simulations for \textcolor{black}{several disorder levels. Figure \ref{fig:kfl} summarizes the SC temperature \tc against the $k_Fl$ estimated in NbN samples \cite{Chand2012,Cheng2016}. Since we used the NbN samples with \tc~=~14.9~K, 12.5~K, and 9.1~K, denoted by the horizontal gray dashed lines in Fig. \ref{fig:kfl}, we chose $k_Fl$ = 3, 6, and 10 (indicated by the vertical black arrows in Fig. \ref{fig:kfl}) in the numerical simulations. To further confirm the robustness of the amplitude-mode predominance at $\Omega=2\Delta$, we also performed the numerical simulation for $k_Fl=30$.}

We present the numerical results \textcolor{black}{of the TH and FH nonlinear current $amplitudes$ for different $k_Fl$ in Figs.~\ref{fig:sm}(a-d) and (e-h), respectively. The corresponding TH and FH nonlinear current $intensities$ for $k_Fl = 10$ are shown in Figs. 4(a) and (b) of the main text, respectively.}  We find that (i) the nonlinear FH spectra hosts peaks at $\Omega=\Delta$ and $\Omega=2\Delta$ and (ii) the TH contribution is peaked at $\Omega=2\Delta/3$, $\Omega=\Delta$ and $\Omega=2\Delta$.   This is fully consistent with our theoretical expectation of the resonance conditions summarized in Table \ref{tab1}.
Remarkably, the $\Omega=2\Delta$ resonance for both TH and FH responses are order-of-magnitude enhanced by the amplitude mode regardless of the disorder level. While the reason for the dominance of the amplitude mode is still unclear at the present time, these numerical results reveal that the experimentally observed resonance at $\Omega=2\Delta$ in the FH nonlinear signal can be attributed to the amplitude mode in the entire investigated disorder range.
In contrast, the other resonances are strongly influenced by the disorder level. For instance, Fig.~\ref{fig:sm}(a)-(c) show that the amplitude-mode contribution at $\Omega=\Delta$ (as well as the one here discussed at $\Omega=2\Delta$) in the TH increases going from \textcolor{black}{the $k_Fl=30$ case to the more disordered $k_Fl=3$ case,} in agreement with the previous works \cite{Silaev2019,Seibold2021,Tsuji2020}.

In order to compute the spectral features at finite temperatures and for an arbitrary pulse shape, we modeled the nonlinear kernel as the sum of three Lorentzian functions as
\begin{align}
\label{eq:kt}
&K^{para}(\omega_1,\omega_1+\omega_2,\omega_1+\omega_2+\omega_3) \nonumber \\
&=\alpha^{(1)}L(\omega_1)+\alpha^{(2)}L(\omega_1+\omega_2)+\alpha^{(3)}L(\omega_1+\omega_2+\omega_3),
\end{align}
where $\alpha^{(i)}$ is the weight for each contribution and $L(\omega)=1/[(\omega+i\delta)^2-4\Delta^2]$ is a Lorentzian function peaked at $2\Delta$, i.e.\ the characteristic energy scale of both BCS and amplitude-mode contributions.

\begin{figure}
\includegraphics[width=\columnwidth]{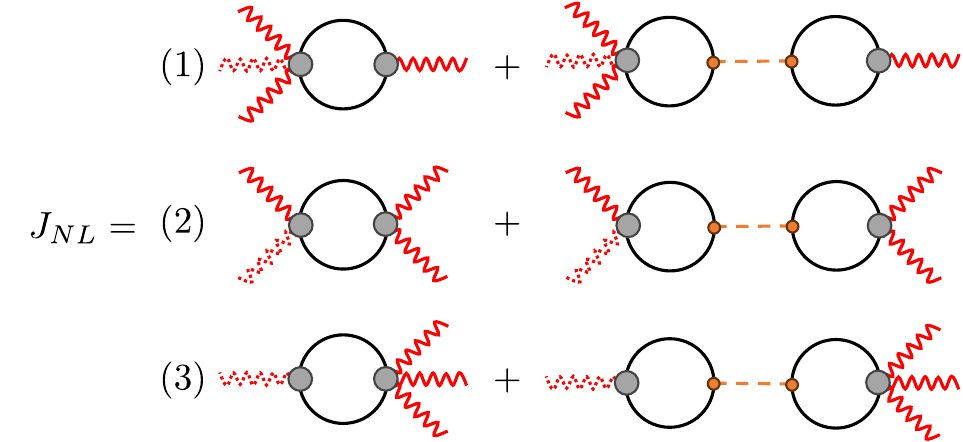}
\caption{Diagrams contributing to the nonlinear kernel (\ref{eq:kt}), with each row weighted by the corresponding $\alpha^{(i)}$. In the model, we effectively approximate both BCS and amplitude-mode contributions (first and second columns, respectively) as Lorentzian propagators at energy $2\Delta$. The middle row is explicitly the diamagnetic contributions, whereas the other rows are paramagnetic ones.}
\label{fig:diagrams}
\end{figure}

\begin{figure}[t]
\includegraphics[width=\columnwidth]{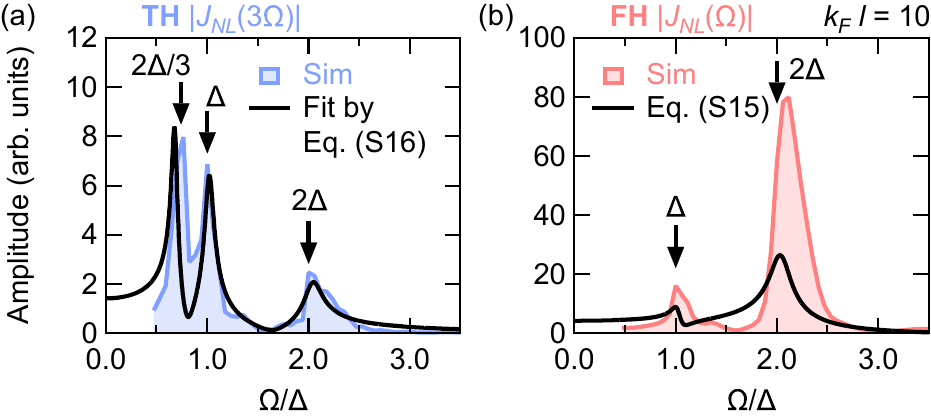}
\caption{ (a) Fitting to the numerical simulated amplitude of the TH nonlinear response as a function of the driving frequency $\Omega$ using Eq. (\ref{eq:mod3}). The  (b) The FH nonlinear response computed by Eq. (\ref{eq:mod1}) with the weights $\alpha^{(i)}$ obtained from the fitting in (a). Here, the disorder level is $k_Fl=\textcolor{black}{10}$.}
\label{fig:fit}
\end{figure}

The kernel in Eq.~(\ref{eq:kt}) approximates the possible diagrammatic contributions to the nonlinear current shown in Fig.~\ref{fig:diagrams}, which is the minimal model to account for the numerically observed positions of the resonances also foreseen in Ref.\ \cite{Benfatto2023}. \textcolor{black}{It should be stressed that the diagrams of Fig.\ \ref{fig:diagrams} must be interpreted as a mapping of the real paramagnetic diagram of Fig.~\ref{fig:paradia} into the effective Kubo-like diagrams of Fig.~\ref{fig:diagrams}, leading each one to a specific resonant combination of the incoming frequencies. Such a structure is exemplified in Eq.\ (\ref{eq:kt})}. 
\textcolor{black}{By plugging Eq. (\ref{eq:kt}) into the general expression (\ref{eq:jpara}) for monochromatic fields centered at $\Omega$, the FH nonlinear current can be written as}
\begin{align}
\label{eq:mod1}
J_{NL}(\Omega)&=\alpha^{(1)}[2L(\Omega)+L(-\Omega)] \nonumber \\
&+\alpha^{(2)}[2L(0)+L(2\Omega)]+3\alpha^{(3)}L(\Omega).
\end{align}
Here, we take into account the combination prefactors. Similarly, we can express the TH nonlinear current as
\begin{align}
\label{eq:mod3}
J_{NL}(3\Omega)=\alpha^{(1)}L(\Omega)+\alpha^{(2)}L(2\Omega)+\alpha^{(3)}L(3\Omega).
\end{align}
\textcolor{black}{We determined the weights $\alpha^{(i)}$ in Eqs. \eqref{eq:mod1} and \eqref{eq:mod3} by fitting the exact numerical results for the BCS and amplitude-mode contributions ($k_Fl=6$) with Eq. \eqref{eq:mod3}.} In Fig.~ \ref{fig:fit}(a), we compare the numerical results for the TH nonlinear current with the fit using Eq. \eqref{eq:mod3}.  Figure~\ref{fig:fit}(b) shows the FH nonlinear current calculated by Eq. \eqref{eq:mod1} with the same $\alpha^{(i)}$ fitted on the TH, compared with the corresponding numerical simulation. Despite the simplification of the model in Eq. \eqref{eq:kt}, Eqs. \eqref{eq:mod1} and \eqref{eq:mod3} qualitatively reproduces the numerical results for both the FH and TH.

\textcolor{black}{Next, we expand this model to the case of an incoming pulse with finite spectral bandwidth. By plugging (\ref{eq:kt}) into the general formula Eq.~(\ref{eq:jpara}) for the nonlinear current, we can write the nonlinear current as the sum of three contributions as $J_{NL}(\omega)=J_{NL}^{(1)}(\omega)+J_{NL}^{(2)}(\omega)+J_{NL}^{(3)}(\omega)$, whose corresponding diagrams are shown in Fig.~\ref{fig:diagrams}.} Here, we approximate the frequency dependence of $K^{para}$ in Fig.~\ref{fig:paradia} as the sum of three Kubo-like processes, where the incoming photon lines are assigned to each effective, frequency-independent vertex in such a way that the resonance of the fermionic propagator is at the position foreseen by the numerical simulation. \textcolor{black}{For instance, the paramagnetic processes which allows one to insert single-photon lines in the diagrams, is approximated as the effective odd-photon vertices of the first and third lines of Fig.~\ref{fig:diagrams}.} The perturbative computation of the actual diagram in Eq.~\eqref{eq:jpara}, which necessarily involves the inclusion of disorder in some approximation schemes, has indeed confirmed the appearance of a peak at $\Omega=2\Delta/3$ in the TH at moderate disorder \cite{Tsuji2020,Silaev2019}. With this in mind, the three processes lead to 
\begin{equation}
J_{NL}^{(1)}(\omega)=3\alpha^{(1)}\int \text{d}\omega^{\prime} A(\omega^{\prime})A^{(2)}(\omega-\omega^{\prime})L(\omega^{\prime}),
\end{equation}
\begin{equation}
J_{NL}^{(2)}(\omega)=3\alpha^{(2)}\int \text{d}\omega^{\prime} A(\omega-\omega^{\prime})A^{(2)}(\omega^{\prime})L(\omega^{\prime}),
\end{equation}
\begin{equation}
\label{eq:3rd}
J_{NL}^{(3)}(\omega)=\alpha^{(3)}L(\omega)\int \text{d}\omega^{\prime}\, A(\omega^{\prime})A^{(2)}(\omega-\omega^{\prime}).
\end{equation}
We find that $J_{NL}^{(2)}(\omega)$ has the same frequency dependence as diamagnetic processes, which have an intrinsic Kubo-like nature. Finally, in order to include the finite temperature effect, we multiply the Lorentzian by the factor $\Delta(T)^2$, and we shift its peak with temperature $T$ as
\begin{equation}
L(\omega,T)=\frac{\Delta(T)^2}{(\omega+i\delta)^2-4\Delta(T)^2}.
\end{equation}
In this way, the kernel softens approaching $T_c$ and mimics the temperature evolution of the nonlinear response in the clean case \cite{Cea2016}. 

\textcolor{black}{Firstly, we calculated the nonlinear current driven by the narrow-band THz pulses. We employed} an incoming pulse with a Gaussian spectrum $E(t)=E_0\exp({-\frac{t^2}{2\tau^2}})\cos(\Omega t)$ with {$\textcolor{black}{\Omega/2\pi}=0.63$~THz, the frequency of the experimental band-pass filter, and $\tau\sim3$~ps. We modeled the scaling in temperature of the SC gap amplitude as in the experiments, with $\textcolor{black}{2\Delta(0)/2\pi}=1.2$~THz fitted from the optical conductivity at 5~K. Following Ref.~\cite{Cea2016}, in the transmission geometry, the TH and FH intensities are proportional to $|J_{NL}(\Omega)|^2$ and $|J_{NL}(3\Omega)|^2$ respectively, evaluated through Eq.~\ref{eq:kt} keeping a finite broadening $\textcolor{black}{\delta/2\pi}=0.12$~THz. We stress that since the $\alpha^{(i)}$ are kept constant, the relative height between the FH and TH intensities is fixed by our model and thus can be directly compared with the corresponding quantity in the experiments, showing a very good agreement as shown in Fig.~4(c) in the main text.

Secondly, we computed the nonlinear current driven by broad-band THz pulses at $T=0$ using the incoming pulse in the experiment. The resulting nonlinear current is presented in Fig.~4(d) and has three peaks highlighted by vertical bars at $\omega=\Omega$, $2\Delta$, and $3\Omega$, qualitatively agreeing with the experimental nonlinear signal plotted by the purple curve in Fig.~1(d). 

\textcolor{black}{
\section{Analysing the THz 2DCS spectra}}
\textcolor{black}{
\subsection{Narrow-band THz pulses}
Here we show how one can disentangle the various nonlinear optical processes that contribute to the 2DCS response. The advantage of 2DCS is that multiple excitation processes are generally possible, each of which can give different information about excitations and their lifetimes. For simplicity, we consider two narrow-band THz pulses at frequency $\omega_A = \omega_B = \Omega$. We call the THz pulse swept along the time $\tau$ the A-pulse, and the other pulse as the B-pulse, following the notation in Ref. \cite{Mahmood2021}. In this definition, the A and B pulses give pump and probe photons respectively, in the usual terminology of pump-probe experiments. The different possible processes arise from the different combinations of the three $E$-field interactions required for a $\chi^{(3)}$ response from pulse A or B and the possibilities of the fields $E(\omega)$ or the complex conjugates $E(\omega)^*$ interacting with the sample. The different contributions can be easily mapped onto the  2DCS spectrum in the space of ($\omega_t, \omega_\tau$) using the so-called frequency-vector scheme \cite{Woerner2013,Mahmood2021}, where $\overrightarrow{\omega_A} = \overrightarrow{\omega_{\tau}} + \overrightarrow{\omega_t} $ and $\overrightarrow{\omega_B} = \overrightarrow{\omega_t} $. Since this spectroscopy only monitors the dynamics $\tau\geq0$ when the pump photon arrives before the probe one, we focus on the frequency-vectors for the 2D spectrum for $\tau\geq0$, i.e., the AB pulse sequence, as depicted in Fig.~\ref{fig:freq_vec}. Here we only consider third-order processes, but the scheme is more general~\cite{Woerner2013}.}

\begin{figure}[t]
	\centering
	\includegraphics[width=8cm]{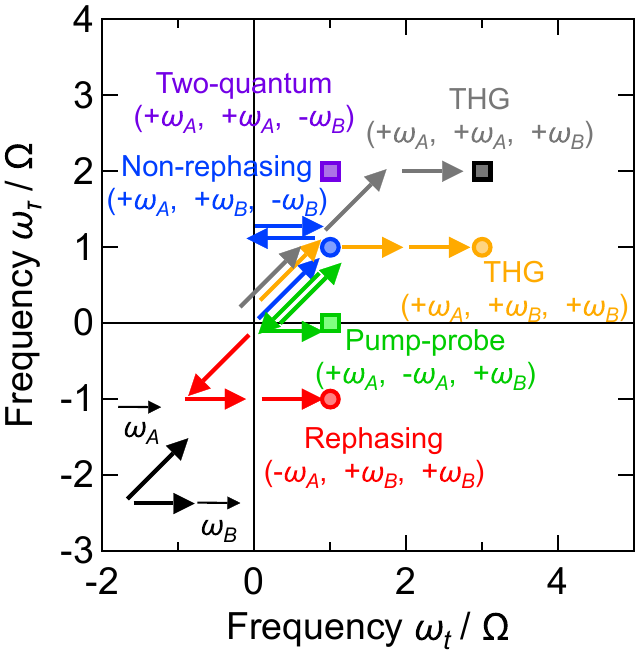}
	\caption{\textcolor{black}{Frequency-vector scheme for the AB pulse sequence. $\overrightarrow{\omega_A}$ is a diagonal vector and $\overrightarrow{\omega_B}$ is a horizontal vector. Conjugated electric field pulses are represented by reversed vectors.}}
	\label{fig:freq_vec}
\end{figure}

\begin{figure}[t]
	\centering
	\includegraphics[width=\columnwidth]{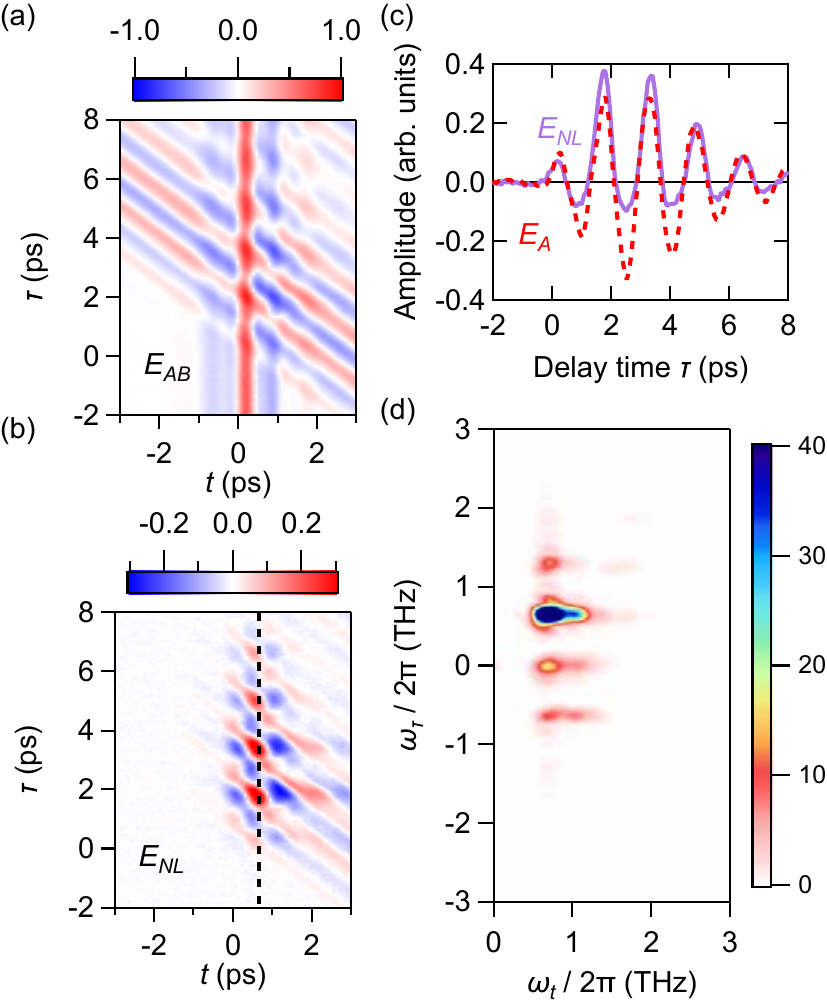}
	\caption{\textcolor{black}{(a) Time traces of the narrow-band A and broad-band B pulses together ($E_{AB}(t,\tau)$) transmitted through the NbN sample (\tc~=~14.9 K) at 5 K~as a function of the sampling time $t$ and the delay time $\tau$. (b) The same plot as (a) but for the nonlinear signal ($E_{NL}(t,\tau)$). (c) Nonlinear signal as a function of $\tau$ measured at $t$~=~0.66~ps (denoted by dashed vertical line in (b)). The red dashed curve shows the time trace of the A-pulse. (d) 2D power spectra of the nonlinear signal measured in (c) for $\tau\geq0$~ps (the AB pulse sequence).}}
	\label{fig:fft_two_add}
\end{figure}

\textcolor{black}{
To confirm this difference, we performed THz 2DCS for the NbN sample (\tc~=~14.9~K) at 5~K using narrow-band (centered at \textcolor{black}{$\Omega/2\pi$}~=~0.63~THz) and broad-band THz pulses for A and B pulses, respectively. This is essentially the same situation as conventional pump-probe experiments in Refs. \cite{Matsunaga2014,matsunaga2015higgs}, except for the slightly more intense probe B $E$-field of 3~\kv. Figures~\ref{fig:fft_two_add}(a) and (b) show the time traces of the A and B pulses together ($E_{AB}(t,\tau)$) and the nonlinear signal ($E_{NL}(t,\tau)$) transmitted after the sample as a function of the sampling time $t$ and the delay time $\tau$. In Fig.~\ref{fig:fft_two_add}(c), we present the nonlinear signal as a function of $\tau$ with the sampling time fixed at $t$~=~0.66~ps together with the narrow-band A-pulse. Importantly, the nonlinear signal oscillates following the A-pulse, indicating the presence of the $\Omega$-oscillatory component which is much larger than the long-lived relaxation component. This is clearly seen in the 2D power spectra of the nonlinear signal for $\tau\geq0$~ps displayed in Fig.~\ref{fig:fft_two_add}(d). One can find peaks at $\omega_{\tau}$~=~$2\Omega$, $\Omega$, 0, and $-\Omega$, corresponding to the two-quantum, non-rephasing, pump-probe, and rephasing responses, respectively, as expected from the general consideration of the frequency-vector scheme in Fig.~\ref{fig:freq_vec}.  A weak signature of the 3$\Omega$ signal can be seen at ($3\Omega,\Omega)$ in the 2D spectra (weaker because of the broad-band probe). The features at $\omega_{\tau}=3\Omega$ are clearly seen in the 2D spectra using two narrow-band THz pulses, as shown in Fig.~\ref{fig:narrow_2d}. Here, the Fourier transform for $\tau$ is performed only for $\tau\geq0$~ps. Figure~\ref{fig:narrow_2d} shows an excellent agreement with the frequency vector scheme in Fig.~\ref{fig:freq_vec}.}

\begin{figure}[t]
	\centering
	\includegraphics[width=4.5cm]{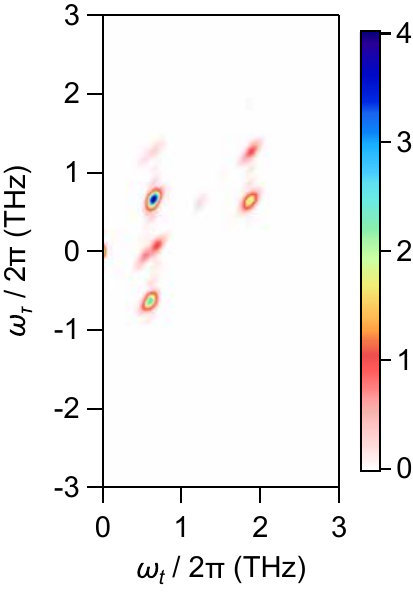}
	\caption{\textcolor{black}{2D power spectra of the nonlinear signal measured with the narrow-band THz pair for $\tau\geq0$~ps (the AB pulse sequence).}}
	\label{fig:narrow_2d}
\end{figure}

\textcolor{black}{
This unambiguously demonstrates that the THz 2DCS can address the rephasing and non-rephasing signals inaccessible by the previously pump-probe spectroscopy.}

\subsection{Broad-band THz pulses}
In this subsection, we analyse the 2D spectra of the nonlinear signal when two THz pulses are broad-band using the theoretical framework in Section VI. We compute the THz 2DCS results with broad-band THz pulses assuming two third-order nonlinear processes: (i) a coherent part associated with an effective nonlinear kernel $K_{coh}$ due to quasiparticle and amplitude mode \textcolor{black}{when two THz pulses are overlapping in time} and (ii) an incoherent response due to quasiparticle recombination \textcolor{black}{when two THz pulses are separated in time}. We model the coherent THz-pair-driven nonlinear current as 
\begin{align}
	\label {eq1m}
	\delta E_{AB}(t,\tau) = \ &\alpha_{coh}\int_{-\infty}^{\infty}\mathrm{d}t^{\prime}K_{coh}(t-t^{\prime})\nonumber \\ 
	&[E_A(t^{\prime},\tau)^2E_B(t^{\prime})+E_A(t^{\prime},\tau)E_B(t^{\prime})^2].
\end{align}
Here, $\alpha_{coh}$ is a constant to match the simulation result with the experimental observation. Since the coherent process displays the resonant behavior with \textcolor{black}{twice the SC gap energy \tdelta} as shown in Fig.~3(c), the Kernel can be assumed to be a Lorentz oscillator centered at \tdelta and expressed in the time-domain as
\begin{align}
	\label {eq2m}
	K_{coh}(t) = \Delta^2\cos(2\Delta t-\phi)\exp(-\gamma t)\Theta(t),
\end{align}
where $\phi$ is the phase of the oscillation and is set to $\pi/4$ to reproduce the data. $\gamma$ is the damping rate of the oscillation and set to 2~THz. $\Theta(t)$ denotes the step function. We use \textcolor{black}{$2\Delta/2\pi$}~=~1.2~THz at 5~K.

\begin{figure}[t]
	\centering
	\includegraphics[width=\columnwidth]{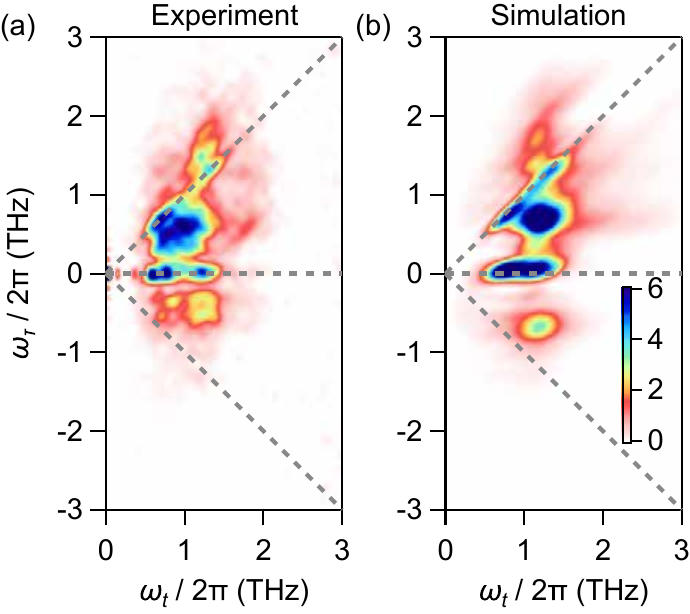}
	\caption{(a) 2D power spectra of the nonlinear signal measured at 5~K with the broad-band THz pulses. (b) The same plot as (a) but in the simulation.}
	\label{fig:fft_two_d}
\end{figure}

\begin{figure*}[t]
	\centering
	\includegraphics[width=\textwidth]{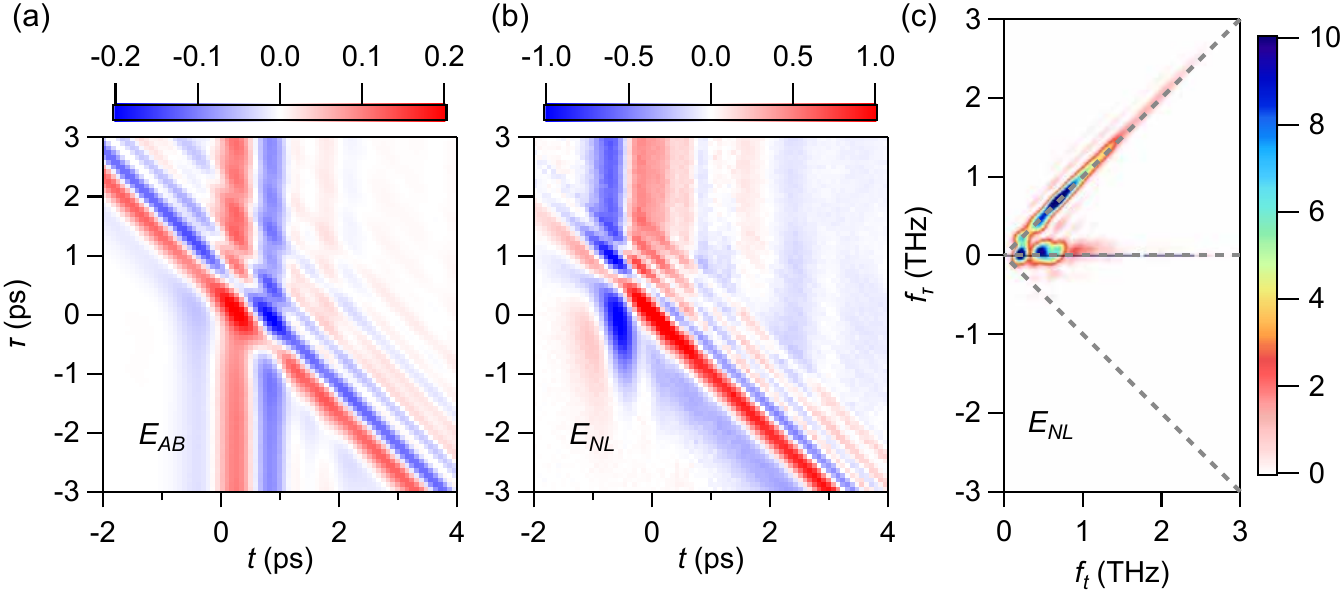}
	\caption{\textcolor{black}{The THz 2D data for the A and B peak $E$-fields of 72 and 82~kV/cm. (a) Time traces of the A and B pulses together ($E_{AB}(t, \tau)$) transmitted after the NbN sample (\tc~=~14.9~K) at 5~K as a function of the sampling time $t$ and the delay time $\tau$. (b) The same plot as (a) but for the nonlinear signal $E_{NL}(t, \tau)$. (c) 2D power spectra of the nonlinear signal shown in (b).}}
	\label{fig:thz_2d_strong}
\end{figure*}

Next, we formulate the incoherent signal for $\tau\geq0$~ps, where the A-pulse arrives earlier than the B-pulse, using the following expression for the usual pump-probe response as \cite{Udina2019,Giorgianni2019},
\begin{align}
	\label {eq3m}
	\delta E_B(t,\tau)=\alpha_{inc}E_B(t)\int_{-\infty}^{\infty}\mathrm{d}t^{\prime}K_{inc}(t-t^{\prime})E_{A}(t^{\prime},\tau)^2.
\end{align}
Here, $\alpha_{inc}$ is a constant, and $K_{inc}$ models the incoherent response due to quasiparticle recombination. To model the long-lived quasiparticle excitation as experimentally verified in the previous study \cite{Matsunaga2012}, we assume the kernel as $K_{inc}(t)=\Theta(t)\exp(-\Gamma t)$, where $\Gamma$ is the damping rate of the quasiparticle excitation and set to 0.5~THz. The incoherent response for $\tau<0$~ps is calculated by switching $E_A(t,\tau)$ and $E_B(t)$ in Eq. (\ref{eq3m}). 

Now, we compare the 2D power spectra of the nonlinear signal $E_{NL}(t,\tau)$ in the experiment (Fig.~\ref{fig:fft_two_d}(a)) and the simulation combining $\delta E_{AB}(t,\tau)$, $\delta E_B(t,\tau)$, and $\delta E_A(t,\tau)$ (Fig.~\ref{fig:fft_two_d}(b)). While the intensities of some peaks are different, the positions and shapes of the peaks are reasonably reproduced by this simple simulation, \textcolor{black}{confirming the internal consistency of the theory and experiment}.

\textcolor{black}{Finally, we present the THz 2D spectra measured with the maximum A and B peak $E$-fields of 72 and 82~kV/cm. Figures \ref{fig:thz_2d_strong}(a) and (b) show $E_{AB}(t, \tau)$ and $E_{NL}(t, \tau)$ measured with the GaP detection crystal. Compared to Fig.~1(d) when A and B peak $E$-fields are 3~kV/cm, $E_{NL}(t, \tau)$ for stronger THz pair in Fig.~\ref{fig:thz_2d_strong}(b) displays prominent pump-probe signals, one along the $\tau$ direction for $\tau \geq 0$~ps and the other along the anti-diagonal direction for $\tau < 0$ ps. Accordingly, the power spectrum of $E_{NL}(t, \tau)$ in Fig.~\ref{fig:thz_2d_strong}(c) displays streaks along $f_\tau=0$ and  $f_\tau=f_t$.}

\bibliographystyle{apsrev4-1}
\bibliography{NbN.bib}